\documentclass[pra,twocolumn,preprintnumbers,amsmath,amssymb,superscriptaddress,showpacs,longbibliography]{revtex4-1}
\usepackage{graphicx}
\usepackage{amsmath}
\usepackage{graphics}
\usepackage{amssymb}
\usepackage{amsfonts,epsfig}
\usepackage{dsfont}
\usepackage{natbib}
\usepackage{hyperref}
\usepackage{enumitem}
\usepackage{physics}
\usepackage{color}
\usepackage[mathscr]{eucal}
\usepackage[caption=false]{subfig}

\allowdisplaybreaks

\newcommand{\beq}{\begin{equation}}
\newcommand{\eeq}{\end{equation}}
\newcommand{\bea}{\begin{eqnarray}}
\newcommand{\eea}{\end{eqnarray}}
\newcommand{\nn}{\nonumber}
\newcommand{\mean}[1]{\langle{#1}\rangle{}}

\newcommand{\cv}[1]{\left(#1\right)}

\newcommand{\aveb}[1]{\left\langle #1\right\rangle}

\begin{document}

\title{Spectroscopy of classical environmental noise with a qubit subjected to projective measurements}
\author{Fattah Sakuldee}
\email{sakuldee@ifpan.edu.pl}
\affiliation{Institute of Physics, Polish Academy of Sciences, Aleja~Lotnik{\'o}w 32/46, 02-668 Warsaw, Poland}
\author{\L{}ukasz Cywi\'n{}ski}\email{lcyw@ifpan.edu.pl}
\affiliation{Institute of Physics, Polish Academy of Sciences, Aleja~Lotnik{\'o}w 32/46, 02-668 Warsaw, Poland}

\begin{abstract}
We show theoretically how a correlation of multiple measurements on a qubit undergoing pure dephasing can be expressed as environmental noise filtering. The measurement of such correlations can be used for environmental noise spectroscopy, and the family of noise filters achievable in such a setting is broader than the one achievable with a standard approach, in which dynamical decoupling sequences are used. We illustrate the advantages of this approach by considering the case of noise spectrum with sharp features at very low frequencies. We also show how appropriately chosen correlations of a few measurements can detect the non-Gaussian character of certain environmental noises, particularly the noise affecting the qubit at the so-called optimal working point. 
\end{abstract}

\date{\today}
		
\maketitle

\section{Introduction}
When a qubit experiences pure dephasing due to coupling to classical Gaussian noise, measurement of its coherence decay under application of an appropriately chosen dynamical decoupling (DD) sequence of unitary operations can be used to reconstruct the power spectral density of the environmental noise \cite{Degen_RMP17,Szankowski_JPCM17}. It is also possible to extend this approach to reconstruction of polyspectra of non-Gaussian noise \cite{Norris_PRL16,Sung2019}.
While this DD-based noise spectroscopy method has found widespread experimental application to multiple kinds of qubits \cite{Degen_RMP17,Szankowski_JPCM17,Almog_JPB11,Biercuk_JPB11,Bylander_NP11,Alvarez_PRL11,Kotler_Nature11,Medford_PRL12,Staudacher_Science13,Muhonen_NN14,Romach_PRL15,Malinowski_PRL17}, using sequences of many single-qubit operations is not without drawbacks:Finite duration and imperfect fidelity of these operations, and also the fact that the qubit is continuously exposed to the noise during the application of the sequence, all limit the range of frequency that is accessible with this method. 
Development of qubit-based environmental noise spectroscopy methods avoiding the use of $\pi$ pulses is thus, apart from being simply theoretically interesting \cite{Liu_NJP10,Wang_PRL19,Ma_PRA18,Muller_SR16}, also of  practical importance \cite{Fink_PRL13,Bechtold_PRL16,Gefen_PRA18,Pfender_NC19,Cujia_Nature19,Do2019}. 

Using correlations between the results of two time-delayed projective measurements on a qubit, in order to obtain information on the environmental noise correlation function or spectral density, was discussed a few years ago \cite{Young_PRA12,Fink_PRL13}. In this paper, we construct a general framework for the description of correlations of multiple projective measurements on a qubit subjected to pure dephasing due to external classical noise (of Gaussian or non-Gaussian character). The main result is casting the expressions for correlators of multiple measurements into a form that clearly shows how their expectation values are connected to noise filtering. In this way we establish a direct analogy of measurement-only protocols with all the research done so far on noise spectroscopy by dynamical decoupling. We also give examples of noise spectra for which the use of measurement-based protocol can lead to higher accuracy and sensitivity of reconstruction of their certain features, compared to the DD-based protocol. Finally, we show how a correlation of three measurements on a qubit can be used to provide evidence of the non-Gaussian character of the environmental noise. 

Note that in another paper \cite{Sakuldee_operations} we discuss the general relationship between applying a DD sequence to a qubit and performing a sequence of projective measurements on it. In that work we make no assumptions about the nature of the environment,~if its fluctuations are classical or quantum, and the qubit-environment coupling. The basic relation between the correlators of multiple measurements and decoherence signal obtained in a DD experiment that holds in the case of the environment being a source of classical noise (and which underpins all the results given in this paper), can be viewed as a particular application of the results from \cite{Sakuldee_operations}. However, due to apparent widespread applicability of the classical noise model of the environment (see \cite{Degen_RMP17,Szankowski_JPCM17} and references therein), and in order to keep the paper self-contained, we have derived this relation here using a simple approach applicable in the case of interest. 

The paper is organized in the following way. In Section \ref{sec:general} we give a general theory for correlation of projective measurements on a qubit that undergoes pure dephasing due to external classical noise. In particular, we show how the expectation values of appropriate linear combinations of correlation functions can be expressed as averages over relative phases of the qubit states that are given by integrals over noise multiplied by a piecewise-constant modulation functions, taking on values of $\pm~1$ and $0$. For such a qubit subjected to dynamical decoupling with short $\pi$ pulses, an analogous picture holds, only with filter functions taking on only $\pm~1$ values.
Then, in Sec.~\ref{sec:Gaussian} we focus on the case of Gaussian noise, for which we can write closed formulas for correlation functions in terms of overlaps between the power spectrum of the noise and frequency-domain filter functions. We discuss there how our theory generalizes the results of \cite{Fink_PRL13} to the case of multiple measurements. Most importantly, we analyze the advantages that the measurement-based noise spectroscopy has over the DD-based one for a particular case of spectrum: one that has very sharp spectral features at low frequency, superimposed over a broadband background of lower power. 
In Sec. \ref{sec:related} we discuss the relation between the measurement-correlation protocols discussed previously and other protocols. In Sec.~\ref{sec:Bechtold} we explain the connection between our results and those of the experiment from \cite{Bechtold_PRL16}, in which a multiple-measurement scheme similar to the one discussed here was used, and in Sec.~\ref{sec:Meriles} we compare our results to the ones obtained using a protocol proposed in \cite{Laraoui_NC13}, where an echo-like sequence involving both $\pi$ and $\pi/2$ pulses was employed. In Sec.~\ref{sec:noreinit} we show how all the previous results can be applied to the protocol in which the qubit is not reinitialized after each measurement.
Finally, in Sec.~\ref{sec:nonGaussian} we discuss how correlations of appropriately chosen measurements can be used to witness the non-Gaussian character of a subclass of non-Gaussian environmental noises.

%%%%%%%%%%%%%%%%
%%% SINGLE QUBIT
%%%%%%%%%%%%%%%%
\section{General theory} \label{sec:general}
We focus on a single qubit that experiences pure dephasing due to coupling to an environment that is a source of classical noise $\xi(t)$. The Hamiltonian of the system is given by
\beq
\hat{H}(t) = \frac{1}{2}\left[ \Omega + \xi(t)  \right] \hat{\sigma}_{z} \,\, ,
\eeq
where $\Omega$ is the energy splitting of the qubit. We furthermore assume now that the noise has zero mean $\mean{\xi(t)} \! =\! 0$ (or that this mean is included in the observable qubit splitting $\Omega$) and that it is stationary. Importantly, initially, we {\it do not} assume that the noise is Gaussian. 

In the following we will consider projective measurements of the qubit performed along both $x$ and $y$ axes. In the presence of finite $\Omega$ they should be understood as being performed in the laboratory frame, i.e.,~along fixed axes. This is a natural setup for certain  qubits, e.g.~a singlet-triplet spin qubit based on a double quantum dot \cite{Taylor_PRB07,Medford_PRL12,Malinowski_NN17,Malinowski_PRL17}.
A more commonly encountered case of a qubit controlled and measured in the rotating frame, when unitary operations are performed by AC pulses of transverse fields resonant with qubit's energy splitting $\Omega,$ corresponds to setting $\Omega\! =\! 0$ below and considering the measurements at various times to be done along $x$ and $y$ axes in the rotating frame. 

\subsection{Correlations of multiple measurements as noise filters}
Following Ref.~\cite{Fink_PRL13} we consider now a protocol in which the qubit is initialized in $\ket{+x}$ state, where $\ket{\pm x} \! =\! \frac{1}{2}(\ket{\uparrow} \pm \ket{\downarrow})$ and $\ket{\uparrow}$ and $\ket{\downarrow}$ are eigenstates of $\hat{\sigma}_z$. After initialization at time $t \! =\!  0$, the qubit precesses under the influence of noise for time $\tau_1$, and then it is subjected to projective measurement in an eigenbasis of either the $\hat{\sigma}_x$ or $\hat{\sigma}_y$ operator at time $t_{1} \! = \! \tau_1$.  Subsequently, at time $t_{1}+\delta t_1$, where $\delta t_1$ is the waiting time after the first measurement, the qubit is reinitialized in the state $\ket{+x}$ and it evolves for time $\tau_2,$ after which it is measured again. 

\begin{figure}[tb]
		\hspace{0cm}\includegraphics[scale=1]{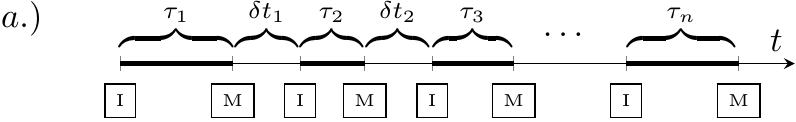}
	\hfill
	\begin{center}
		\includegraphics[scale=1]{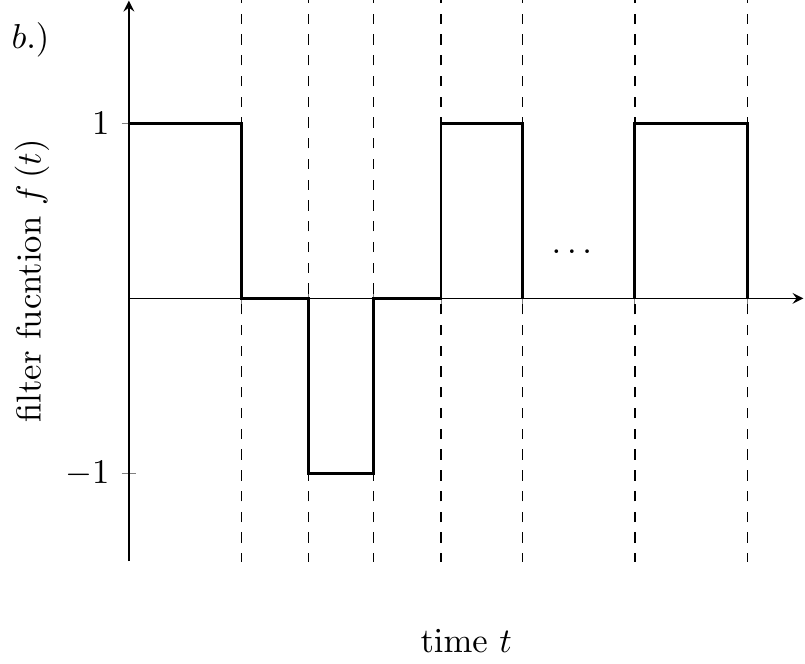}
	\end{center}
	\hfill
	\begin{center}
		\includegraphics[scale=1]{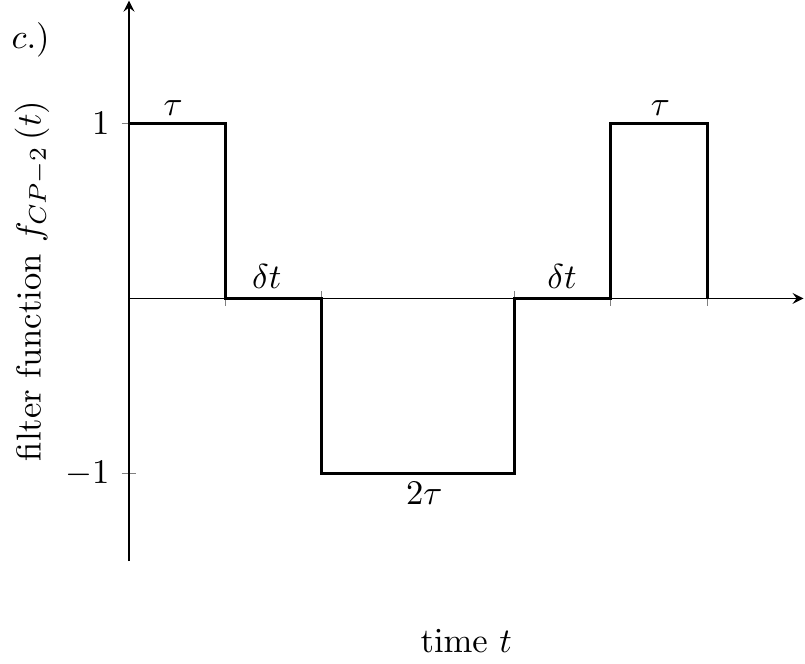}
	\end{center}
	\caption{Examples of sequential measurements protocols. a) Schematics of repetitions of initialization-phase encoding measurement delay. b) One of the time-domain filter functions corresponding to the timing pattern from a) and the $g(1,-1,1,\ldots,1)$ correlation [see Eqs.~(\ref{eq:g}) and (\ref{eq:gf})]. c) Measurement-induced filter generalizing the dynamical decoupling filter corresponding to the two-pulse Carr-Purcell sequence.}\label{fig:single_qubit}
\end{figure}

We now generalize the two-measurement setup from Ref.~\cite{Fink_PRL13} to the case of $n$ measurements. We consider a sequence in which the $k$-th initialization occurs at time $t_{k}-\tau_k$ (with $t_{1}\! =\!\tau_1$), the $k$-th evolution of the qubit interacting with the noise lasts for $\tau_k$, the $k$-th measurement occurs at time $t_{k}$, and the delay between the $k$-th measurement and $k+1$ initialization is $\delta t_k$ [see Fig.~\ref{fig:single_qubit} a)]. 

For a given realization of $\xi(t)$, the state subjected to the $k$-th measurement is
\beq
\ket{\alpha_k} \equiv e^{-i\alpha_k \hat{\sigma}_z /2}\ket{+x} \,\, ,
\eeq
with $\alpha_k = \Omega\tau_k + \Phi_k$, where $\Phi_k$ is the angle of rotation due to the noise,
\beq
\Phi_k \equiv \int_{t_k-\tau_k}^{t_k} \xi(t) \mathrm{d}t \,\, . \label{eq:Phik}
\eeq
The probability of obtaining a $\pm 1$ result when performing measurement of $\hat{\sigma}_{x}$ on state $\ket{\alpha_k}$ is
\beq
p_{x}(\pm|\alpha_k) = \frac{1}{2}(1\pm \cos\alpha_k ) \,\, ,\label{eq:prob_single_x}
\eeq
while the probability of obtaining a $\pm 1$ result when measuring $\hat{\sigma}_{y}$ is
\beq
p_{y}(\pm|\alpha_k) = \frac{1}{2}(1\pm \sin\alpha_k ) \,\, .
\eeq
We consider now the expectation value of the correlation of results of $n$ measurements of $\hat{\sigma}_{x}$ or $\hat{\sigma}_{y}$. We assume that the $n$-measurement protocol is repeated a large number of times, so that averaging over the measurement results corresponds to averaging both over all the possible values of $\alpha_k$ (with $k\! =\! 1, \ldots, n$) and over the results of projective measurements for each $\alpha_k$. With $(a_1,\ldots ,a_n)$ in which $a_{k}\! =\! x$, $y$ denote the measurement axes, and $(m_1,\ldots , m_n),$ in which $m_{k} \! =\! \pm 1$ denote the measurement results, for given $\alpha_1,\ldots,\alpha_n$, the probability of getting the string $(m_1,\ldots ,m_n)$ of results is
\beq
p_{a_1,\ldots,a_n|\alpha_1\,\ldots,\alpha_n}(m_1,\ldots,m_n) = \prod_{k=1}^{n} p_{a_{k}}(m_k|\alpha_k)  \,\, . \label{eq:p_reinit}
\eeq
The correlation function that we are interested in is given by 
\begin{align}
& C_{a_1,\ldots,a_n} (t_1,\tau_1 ; \cdots  ;t_n,\tau_n ) = \sum_{m_1\! =\! \pm 1} \cdots \sum_{m_n\! =\! \pm 1}   \nonumber\\
& \Big\langle p_{a_1,\ldots,a_n}(m_1,\ldots,m_n|\alpha_1,\ldots,\alpha_n) m_1 m_2 \cdots m_n   \Big\rangle_{\Phi_1, \ldots, \Phi_n} \nonumber\\
& = \left\langle \sum_{m_1\! =\! \pm 1} p_{a_1}(m_1|\alpha_1) m_1 \ldots \sum_{m_n\! =\! \pm 1} p_{a_n}(m_n|\alpha_n) m_n   \right\rangle_{\Phi_1, \ldots, \Phi_n} \nonumber\\
& = \big\langle e_{a_1}(\alpha_1) \cdots e_{a_n}(\alpha_n)  \big\rangle_{\Phi_1, \ldots, \Phi_n}  \,\, , \label{eq:C}
\end{align}
where $\langle \cdots \rangle_{\Phi_1, \ldots, \Phi_n}$ denotes averaging over the distribution of phases $\Phi_{k}$ (noise-induced stochastic parts of $\alpha_k$) and the functions $e_{x(y)}(\alpha)$ are expectation values of $\hat{\sigma}_{x(y)}$ on state $\ket{\alpha}$, i.e.,~$e_{x}(\alpha) \! =\! \cos\alpha$ and $e_{y}(\alpha) \! =\! \sin\alpha$.
The correlator for $n$ measurements in the $x$ basis can thus be written as
\beq
C_{x\ldots x}(t_1,\tau_1; \ldots  t_n,\tau_n)  = \mean{\cos\alpha_1 \cos\alpha_2 \ldots \cos\alpha_n}_{\Phi_1 \ldots \Phi_n} \,\, ,  \label{eq:Cxxxx}
\eeq
All the other correlators, corresponding to other choices of measurement axes, are obtained by replacing the respective $\cos \alpha_k$ by $\sin \alpha_k$ whenever $a_k \! =\! y$. 

In order to most easily see the relation between the above correlators and the physical picture of noise filtering, let us focus on measurements of $\hat{\sigma}_{s} \! =\! \hat{\sigma}_{x} +is\hat{\sigma}_{y}$. For a sequence of $n$ measurements defined by a set of measurement times $(t_1,\ldots,t_n)$ and interaction times $(\tau_1,\ldots,\tau_n)$ one should measure all the $2^{n}$ correlators $C_{a_1,\ldots, a_n}(t_1,\tau_1 ; \ldots;  t_n,\tau_n )$ corresponding to all the possible choices of $x$ and $y$ measurement axes, and combine the results to obtain 
\begin{align}
g\cv{s_1,s_2,\ldots,s_n} & = \aveb{\sigma_{+}\cv{t_1}\sigma_{s_2}\cv{t_2}\cdots\sigma_{s_n}\cv{t_n}} \,\, ,\nonumber\\
& =\aveb{\exp\cv{i\displaystyle\sum_{k=1}^n s_k(\Omega\tau_k + \Phi_k)}} \,\, , \label{eq:g}
\end{align}
for a desired set of $(s_1,\ldots,s_n)$ values. (we have set $s_{1}\! =\! 1$ without any loss of generality of the below results).
 
We use now the definition of the $\Phi_k$ phase from Eq.~\eqref{eq:Phik} to arrive at
\beq
g\cv{s_1,s_2,\ldots,s_n} = e^{i \Omega \sum_{k} s_k \tau_k} \left\langle \exp \left( i\int_{0}^{t_{n}} f(t)\xi(t) \mathrm{d}t \right)  \right\rangle_{\xi} \,\, ,\label{eq:gf}
\eeq
in which we recognize the expression wellknown from calculations of dynamical decoupling coherence signals for a qubit coupled to classical noise \cite{deSousa_TAP09,Cywinski_PRB08,Biercuk_JPB11,Szankowski_JPCM17}. In the case considered here, the temporal filter function $f(t)$ is given by
	\begin{equation}
f\cv{t} = \left\{\begin{array}{lr}
	1, & 0<t<\tau_1\\
	0, & t_k<t<t_k+\delta t_k~\forall~k\\
	s_k, & t_k-\tau_k<t<t_k~\forall~k \,\, ,
\end{array}\right. \label{eq:meas_f}
		\end{equation}
with examples of filters shown in Figs.~\ref{fig:single_qubit}b) and \ref{fig:single_qubit}c).

It easy to see that if we set all $\delta t_k$ equal to zero, so that we consider an experiment in which measurements are followed immediately by re-initializations of the qubit [``immediately'' physically means on a timescale on which the noise $\xi(t)$ is too a good approximation constant], and we look at correlators with $s_{k}\! =\! (-1)^{k+1}$, the corresponding filter functions are equal to the ones that appear in the calculation of qubit's coherence after application of a dynamical decoupling sequence of short $\pi$ pulses applied at $t_{k}$ times. For example, we consider the $g(1,-1)$ correlation function with $\tau_1\! =\!\tau_2 \! = \! \tau$ and $\delta t_1\! = \! 0$, which is given by
\begin{align}
g(1,-1) & = C_{xx}(\tau,\tau;2\tau,\tau) + C_{yy}(\tau,\tau;2\tau,\tau) \nonumber\\
& -iC_{xy}(\tau,\tau;2\tau,\tau) + iC_{yx}(\tau,\tau;2\tau,\tau) \,\, , \\
& = \left\langle  \exp \left( i\int_{0}^{\tau} \xi(t) \mathrm{d}t -i\int_{\tau}^{2\tau}  \xi(t) \mathrm{d}t \right) \right\rangle \,\, , \label{eq:echo}
\end{align}
which is exactly the spin echo signal obtained when one measures $\hat{\sigma}_{+}$ at time $2\tau$, after applying a $\pi$ pulse (about either the $x$ or $y$ axis) at time $\tau$. Note that if one measures $\sigma_x$ at $2\tau$, the echo signal corresponds to the real part of the above expression, and taking into account that $C_{a_1,\ldots,a_n}$ are real, we have a simpler expression
\beq
\mean{\hat{\sigma}_{x}(2\tau)}_{\mathrm{echo}} = C_{xx}(\tau,\tau;2\tau,\tau) + C_{yy}(\tau,\tau;2\tau,\tau) 
\eeq

Generally, the result for coherence decay under application of a DD sequence of $n-1$ pulses applied at times $t_k$, $k\! =\! 1,\ldots,n-1$, given by $g(1,\ldots,-(-1)^{n})$, is constructed from correlators of $n$ measurements (assuming all $\delta t_k \! =\! 0$) as
\beq
g\cv{1,\ldots,-(-1)^n} = \sum_{k=0}^n\sum_{\pi_k} i^k\cv{-1}^{r\cv{k}}C_{\pi_k\cv{r_1,\ldots,r_k}}
\eeq
where $\pi_k\cv{r_1,\ldots,r_k}$ denotes the sequence of measurement axes $a_1,\ldots,a_n$ of length $n$ containing $k$ items of $y$ at the orders $r_1,\ldots,r_k,$ in the sequence, e.g., for $n=4,$ $\pi_2\cv{2,4}=xyxy,$ and $r\cv{k}=\displaystyle\sum_{j=1}^k r_j.$

Consider now a three-measurement example, the $g(1,-1,1)$ correlation function with $\tau_1\! =\!\tau_3 \! = \! \tau, \tau_2 \! = \! 2\tau,$ and $\delta t_1\! = \! 0,$ which is given by
\begin{align}
g&(1,-1,1)\\
  	&= C_{xxx}(\tau,\tau;3\tau,2\tau;4\tau,\tau) - C_{yxy}(\tau,\tau;3\tau,2\tau;4\tau,\tau) \nonumber\\
& +C_{yyx}(\tau,\tau;3\tau,2\tau;4\tau,\tau) + C_{xyy}(\tau,\tau;3\tau,2\tau;4\tau,\tau) \nonumber\\
& +iC_{yxx}(\tau,\tau;3\tau,2\tau;4\tau,\tau) + iC_{xxy}(\tau,\tau;3\tau,2\tau;4\tau,\tau)\nonumber\\
& -iC_{xyx}(\tau,\tau;3\tau,2\tau;4\tau,\tau) + iC_{yyy}(\tau,\tau;3\tau,2\tau;4\tau,\tau) \,\, , \\
& = \left\langle  \exp \left( i\int_{0}^{\tau} \xi(t) \mathrm{d}t -i\int_{\tau}^{3\tau}  \xi(t) \mathrm{d}t +i\int_{3\tau}^{4\tau} \xi(t) \mathrm{d}t \right) \right\rangle \,\, . \label{eq:CP-2}
\end{align}
The above filter corresponds to a two-pulse Carr-Purcell sequence (CP-2), shown in Fig.~\ref{fig:single_qubit}c).

The above relationship between the expectation values of a correlations function of $n$ measurements on the qubit, and the coherence signals obtained after subjecting the qubit to dynamical decoupling, is the key result of this paper. In principle, it allows for formally straightforward translation of all that is known about DD-based noise spectroscopy of classical dephasing noise \cite{Szankowski_JPCM17}, be it Gaussian or non-Gaussian \cite{Norris_PRL16,Sung2019}, to the setting in which the qubit is subjected only to projective measurements. However, it must be noted that high-precision applications of noise spectroscopy require using large numbers of pulses \cite{Alvarez_PRL11,Szankowski_PRA18}. Correlators $g(s_1,\ldots,s_n)$ that are directly related to coherences considered in DD-based protocols then have to be constructed from an exponentially large number of $C_{a_1,\ldots,a_n}$ correlators. 

Let us however stress that with nonzero $\delta t_{k}$ delay times between measurements and re-initializations of the qubit, linear combinations of $C_{a_1,\ldots,a_n}$ measurements giving $g(1,\ldots,(-1)^n)$ correspond to measurements of coherence of a qubit subjected to noise filtered through $f(t)$ given in (\ref{eq:meas_f}), and this family of functions is richer than the one that appears when considering dynamical decoupling of the qubit (see, however, REf \cite{Laraoui_NC13} and the discussion in Sec.~\ref{sec:Meriles}).
Due to presence of time periods in which $f(t)\! =\! 0$,  it allows for more flexibility in reconstruction of long-time correlations of $\xi(t)$ (low-frequency noise) (see Sec.~\ref{sec:example}).

%%%%%%%%%%%%%%%%%%%%%%%%%%
%% ALL MEASUREMENS ALONG X
%%%%%%%%%%%%%%%%%%%%%%%%%%
\subsection{Filters for all measurements along the same axis}
While the relationship between noise filtering and correlations of multiple measurements is most direct when we consider $g(1,s_2,\ldots,s_n)$ correlation functions from Eq.~(\ref{eq:gf}), every single $C_{a_1,\ldots,a_n}$ correlation function can also be related to measurements of the qubit's coherence after a certain filtering of noise. Let us focus now on the simplest possible case of $C_{x,\ldots,x}$ correlations. Replacing every $\cos \alpha_k$ by $\frac{1}{2}(e^{i\alpha_k} + e^{-i\alpha_k})$ in Eq.~(\ref{eq:Cxxxx}) we arrive at
\beq
C_{x,\ldots,x}(t_1,\tau_1;\ldots;t_n,\tau_n) = \frac{1}{2^n}\sum_{s_1=\pm 1} \ldots \sum_{s_n=\pm 1} g(s_1,\ldots,s_n)  \,\, , \label{eq:Cx_g}
\eeq
which means that a correlator of $n$ measurements along $x$ is given by the sum of coherence signals corresponding to all the possible filters defined by sets of $s_1,\ldots,s_n$ values, i.e.~filters from Eq.~(\ref{eq:meas_f}) with both values of $s_1\! = \! \pm 1$ taken into account.

In the simplest case of correlation of two consecutive measurements of $\hat{\sigma}_{x}$, i.e.,~for $C_{xx}(t_1,\tau_1; t_2, \tau_2) $ considered
 in \cite{Fink_PRL13}, we obtain
\begin{align}
C_{xx}  & =  \frac{1}{4}\left[ g(1,1) + g(-1,-1) + g(1,-1) + g(-1,1)\right] \,\, , \nonumber\\
& = \frac{1}{2}\mathrm{Re}\left[ g(1,1) + g(1,-1)\right] \,\, .\label{eq:Cxx_general}
\end{align}
The first term above corresponds to coherence of the qubit exposed to noise for time periods $t\in [0,\tau_1]$ and $t\in[t_2-\tau_2,t_2]$, while the second corresponds to coherence of the qubit exposed to the noise for both of these two periods, but with the sign of noise flipped during the second one, i.e.~it corresponds to a generalization of the spin echo signal. For $\delta t_1 \! =\! 0$ and $\tau_1 \! =\! \tau_2 \! = \tau$ the first term is simply the coherence of a qubit freely evolving under the influence of noise for time $2\tau$, while the second one corresponds to the echo signal measured after the same time, with the $\pi$ pulse applied at $\tau$. This is of course the structure obtained in \cite{Fink_PRL13}, but here we have explicitly shown how this arises as a special case of the general result, Eq.~(\ref{eq:Cx_g}).

%%%%%%%%%%%%%%%%%%%%%%%%%%
%% LOW-FREQUENCY NOISE
%%%%%%%%%%%%%%%%%%%%%%%%%%
\subsection{Low frequency noise}  \label{sec:low}
Let us see which types of correlations of measurements are immune to noise at the lowest frequencies. When 
\begin{equation*}
\sum_{k} s_{k}\tau_k = 0  \,\,\, \Leftrightarrow \,\,\, \int f(t)\mathrm{d}t = 0 
\end{equation*}
the trivial phase factor vanishes from Eq.~(\ref{eq:gf}), and the influence of the slowest dynamics of $\xi(t)$ is suppressed. Since the low-frequency noise is typically strong for most of the physical realizations of qubits (especially for solid-state based ones \cite{Paladino_RMP14,Szankowski_JPCM17}, but also for qubits based on ion traps \cite{Monz_PRL11}), correlators corresponding to such ``balanced'' filters are of particular interest, as they are expected to have non-negligible values on timescale much longer than the ones that are sensitive to such noise. 

When the qubit is exposed to strong low-frequency noise, all the terms in Eq.~\eqref{eq:Cx_g} that correspond to imbalanced sequences should decay rather quickly to zero as the total time of exposure of the qubit to noise, $\sum_{k} \tau_k$, increases. If the considered set of $t_1,\ldots,t_n$ in fact allows for construction of balanced $f(t)$ filters, then some - but not all - of $2^{n}$ correlations $g$ contributing to Eq.~\eqref{eq:Cx_g} will correspond to such filters. For simplicity let us discuss the case of even $n$ and all the evolution times $\tau_k$ being equal to $\tau$. Then, the number of balanced $g$ correlations is ${n}\choose{n/2}$, which is approximately equal to$2^{n}$ for large $n$, so in this limit the amplitude of $C_{x,\ldots,x}$ correlation should be close to unity for  $n\tau \gtrsim T_{2}^{*}$. However, for small $n$ only a fraction of the correlation signal is so long-lived, e.g.~for $n\! =\! 4$ only $6$ out of $16$ contributions to $C_{xxxx}$ correspond to balanced sequences. Also, the presence of multiple $g$ correlations contributing to the measured signal, with many of them corresponding to very distinct filters, complicates the conversion of the measured signal into information about the noise.

%%%%%%%%%%%%%%%%%%%%
%%% GAUSSIAN RESULTS
%%%%%%%%%%%%%%%%%%%%
\section{Spectroscopy of Gaussian noise}  \label{sec:Gaussian}
Let us focus now on the oftenencountered in experiments (see \cite{Szankowski_JPCM17} and references therein), and theoretically very simple to consider case of Gaussian noise. The statistics of the stochastic process $\xi(t)$ is then fully determined by its autocorrelation function $C(t_1 - t_2) \equiv \mean{\xi(t_1)\xi(t_2)}$, or equivalently by its power spectral density defined as
\beq
S(\omega) = \int_{-\infty}^{\infty} C(t) e^{i\omega t}\mathrm{d}t \,\, . \label{eq:S}
\eeq

\subsection{General formulation}
When noise $\xi(t)$ is Gaussian, averages over noise realizations can be easily performed. Gaussian statistics of $\xi(t)$ means that phases $\Phi_{k}$ are also Gaussian variables, and
\beq
\mean{ e^{i\sum_{k} s_k \Phi_k } } = \exp\left( -\frac{1}{2}\sum_{k,k'} s_k s_{k'}\mean{\Phi_{k}\Phi_{k'}} \right) \,\, ,
\eeq
Eq.~(\ref{eq:gf}) is then transformed into
\begin{align}
g(1,s_2,\ldots,s_n) & = e^{i\Omega\sum_k s_k \tau_k} \times \nonumber\\
& \!\!\!\!\!\!\!\!\!\!\!\!\!\!\! \exp\left( -\frac{1}{2}\int_{0}^{t_n} \int_{0}^{t_n} f(t_1)f(t_2) \mean{\xi(t_1)\xi(t_2)} \mathrm{d}t_1 \mathrm{d}t_2 \right) \,\, .
\end{align}
Using the definition of the noise autocorrelation function and its power spectral density, $S(\omega),$ from Eq.~(\ref{eq:S}), we arrive at \cite{deSousa_TAP09,Cywinski_PRB08,Szankowski_JPCM17}
\beq
g(1,s_2,\ldots,s_n) =  e^{i\Omega\sum_k s_k \tau_k} e^{-\chi_{1,s_2,\ldots,s_n }} \,\, , \label{eq:gchi} 
\eeq
with 
\beq
\chi_{1,s_2,\ldots,s_n} = \int_{0}^{\infty} S(\omega) |\tilde{f}_{1,s_2,\ldots,s_n}(\omega)|^2  \frac{\mathrm{d}\omega}{2\pi}  \,\, , \label{eq:chi}
\eeq
in which $\tilde{f}(\omega)$ is the Fourier transform of the temporal filter function, i.e.,~it is the filter in the frequency domain.

Note that the above results also hold to a good approximation in a small decoherence limit, in which $\chi \! \ll \! 1$, and in fact we have $g \approx 1 - \chi$. Even if the noise is non-Gaussian, in this weak-coupling limit it is enough to use only the second cumulant of the noise to approximate $\chi$ \cite{Szankowski_JPCM17,Kofman_PRL04,Zwick_PRAPL16}.

\subsection{Example: correlation of two measurements}
For the two-measurement protocol from Ref.~\cite{Fink_PRL13} we have then, using Eq.~(\ref{eq:Cxx_general}), that
\beq
C_{xx}(\tau,\tau; \tau+\delta t, \tau)  = \frac{1}{2}\cos(2\Omega\tau)e^{-\chi_{1,1}} + \frac{1}{2}e^{-\chi_{1,-1}} \,\, ,
\eeq
with
\begin{align}
\chi_{1,1} & = \int_0^{\infty} S(\omega) \frac{8}{\omega^2} \sin^2\frac{\omega\tau}{2}\cos^2\frac{\omega(\tau+\delta t)}{2} \frac{\mathrm{d}\omega}{\pi} \,\, , \label{eq:chip}\\
\chi_{1,-1} & = \int_0^{\infty} S(\omega) \frac{8}{\omega^2} \sin^2\frac{\omega\tau}{2}\sin^2\frac{\omega(\tau+\delta t)}{2}\frac{\mathrm{d}\omega}{\pi} \,\, ,\label{eq:chim}
\end{align}
called in \cite{Fink_PRL13} $\frac{1}{2}\chi_{+}$ and $\frac{1}{2}\chi_{-}$, respectively. As discussed in Sec.~\ref{sec:low}, in the case of noise with most of the power spectrum concentrated at low frequencies, the decay of the $\exp( -\chi_{1,1} )$ term occurs much more quickly than that of the $\exp(-\chi_{1,-1})$ term. Assuming $\delta t \! \gg \! \tau$, we have 
\beq
\chi_{1,1} \approx  2 \tau^{2} \int_{0}^{1/\tau} S(\omega) \frac{\mathrm{d}\omega}{\pi} \approx 2\sigma^2 \tau^2 \equiv \left( \frac{2\tau}{T_{2}^{*}}\right)^2 \,\, ,
\eeq
where we considered up to values  such that most of the total power, given by $\sigma^2 \equiv \int_{0}^{\infty} S(\omega) \mathrm{d}\omega/\pi$, is located at frequencies lower than $1/\tau$. Note that under the same conditions, free induction decay of single-qubit coherence would be given by
\beq
\mean{e^{-i\Phi(0,\tau)}}= e^{-\chi_{\mathrm{FID}}(\tau)} \approx \! e^{-(\tau/T_{2}^{*})^2} \,\, ,
\eeq
where $\chi_{\mathrm{FID}} = \int_{0}^{\infty} \frac{S(\omega)}{\omega^2} 2\sin^2\frac{\omega \tau}{2} \mathrm{d}\omega$, and $T_{2}^{*} \! = \!\sqrt{2}/\sigma$. The decay of the $\exp(-\chi_{1,1})$ term is then the same as for the qubit exposed to low-frequency noise for the total time $2\tau$. 

On the other hand, the echolike term $\exp(-\chi_{1,-1})$ decays more slowly, as thecontribution of frequencies lower than approximately $1/\delta t$ is suppressed in Eq.~(\ref{eq:chim}). Analyzing the $\tau$ and $\delta t$dependence of this part of the signal then allows us to infer certain characteristics of the high-frequency part of the spectrum \cite{Fink_PRL13}, just as analysis of the $\tau$ dependence of echo decay gives qualitative information on high-frequency noise \cite{Cywinski_PRB08,deSousa_TAP09,Dial_PRL13}.

%%%%%%%%%%%%%%
% SPECTROSCOPY
%%%%%%%%%%%%%%
\subsection{Spectroscopic reconstruction of power spectral density} \label{sec:spectroscopy}
The most robust methods of reconstruction of $S(\omega)$ (``noise spectroscopy'') from measurements on the qubit, involve applying many $\pi$ pulses to the qubit \cite{Szankowski_JPCM17,Degen_RMP17}, in order to create a filter $f(t)$ that has a well-defined periodic structure, with its basic block being repeated many times. Frequency filters $\tilde{f}(\omega)$ obtained in this case have narrow-pass character \cite{Alvarez_PRL11,Yuge_PRL11,Bylander_NP11,Szankowski_JPCM17,Degen_RMP17}, and the relationship between the measured signal and the noise power spectral density becomes particularly straightforward (although not completely trivial; see, e.g.~\cite{Alvarez_PRL11} and \cite{Szankowski_PRA18}). Let us investigate then the generalized filters (\ref{eq:meas_f}) in such a setting.

\begin{figure}[tb]
	\begin{center}
		\includegraphics[scale=1]{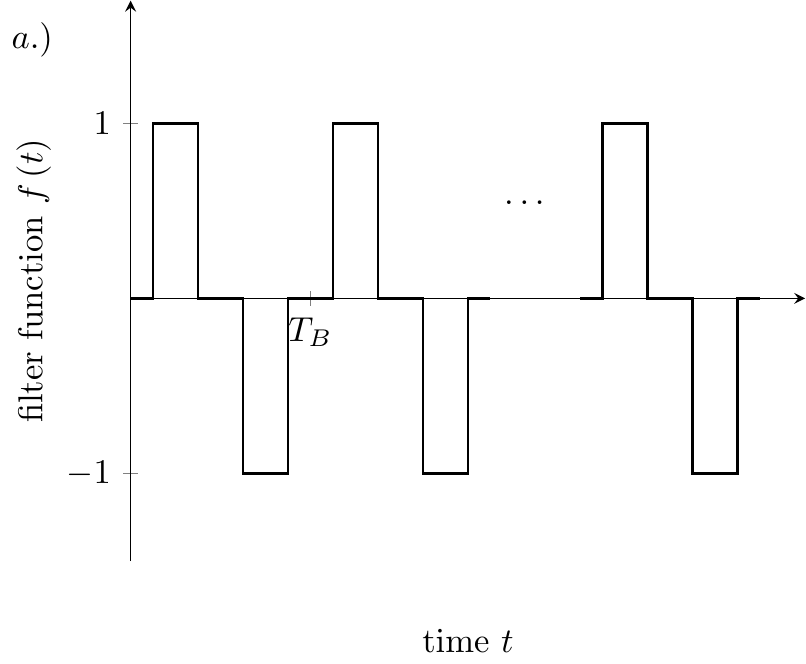}
	\end{center}
	\hfill
	\begin{center}
		\includegraphics[scale=1]{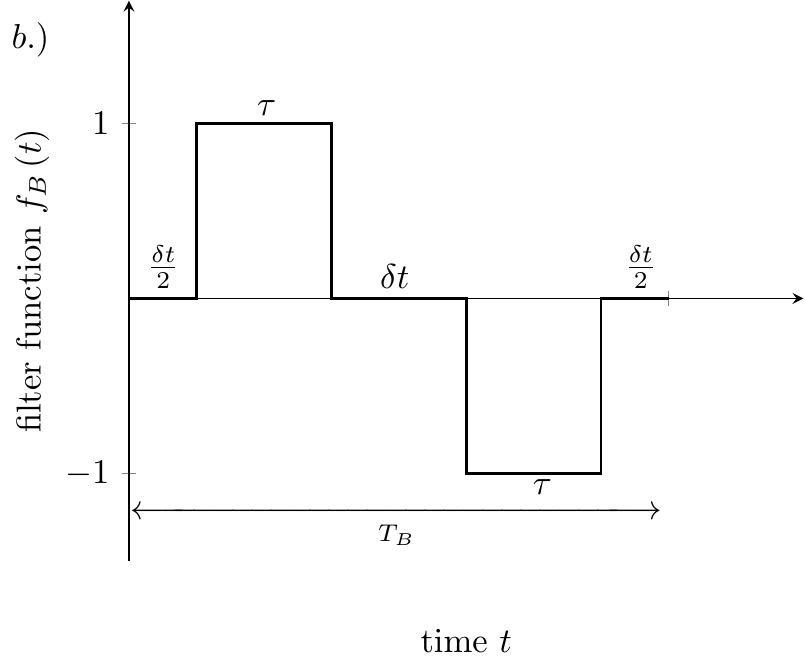}
	\end{center}
	\caption{a) Measurement filter function corresponding to  $g(1,-1,\ldots,-1)$, which consists of $N$ repetitions of the filter block shown in b).  }\label{fig:f_spectr}
\end{figure}

For simplicity, we consider a protocol with an even number of measurements $n\! = \! 2N$, characterized by all $\tau_{k}$ equal to $\tau$ and all $\delta t_k$ equal to $\delta t$. We focus on the $g(1,-1,1\ldots,-1)$ correlation function, corresponding to the $f(t)$ filter shown in Fig.~\ref{fig:f_spectr}a). 
This filter is constructed by repeating a basic block $f_{B}(t)$ of duration $T_{B} \! =\! 2(\tau+\delta t)$shown in Fig.~\ref{fig:f_spectr}b) $N$ times. We can write it as
\beq
f(t) = \Theta(NT_{B} - t) \Theta(t) \sum_{m}c_{m\omega_p} e^{i m \omega_p} \,\, ,
\eeq
where $\omega_p = 2\pi /T_{B}$ is the base frequency of the filter, and the Fourier coefficient $c_{m\omega_p}$ is given by
\beq
c_{m\omega_p} = \frac{1}{T_{B}} \int_{0}^{T_{B}}f_{B}(t) \mathrm{d}t = \frac{2i}{\pi m} \cos \frac{\pi m}{\tau + \delta t}\frac{\delta t}{2} \,\, ,
\eeq
for odd $m$ and zero otherwise. As discussed in Refs.~\cite{Szankowski_JPCM17,Szankowski_PRA18}, for $N\! \gg \! 1$ we can then approximate $|\tilde{f}(\omega)|^2$ as
\beq
\vert\tilde{f}(\omega)|^2 \approx T \sum_m |c_{m\omega_p}\vert^2 T \mathrm{sinc}^2\frac{(\omega -m\omega_p)T}{2} \,\, ,
\eeq
where $T \! =\! N T_{B}$, and the time at which the last measurement is taken is $t_{n} \! =\! T-\frac{\delta t}{2}$, which occurs $T-\delta t$ after the first initialization of the qubit. 

With increasing $N,$ and thus increasing $T$, the terms $T\mathrm{sinc}^2 (\omega -m\omega_p)T/2$ behave more and more as approximations of $\delta$-functions centered at $m\omega_p$, each characterized by a width approximately $2\pi/T$. A sequence of $n \! \gg \! 1$ measurements that lasts for time $T$ corresponds then to a filter $|\tilde{f}(\omega)|^2$ that consists of narrow peaks centered at odd harmonics of $\omega_p$, and the width of band-pass regions decreases as $1/T$. When the filter peaks become sharper than any features of $S(\omega)$, the measured signal is determined by the attenuation function $\chi$ given by
\beq
\chi(T) \approx T \sum_{m>0} |c_{m\omega_p}|^2 S(m\omega_p)  \equiv T R(\omega_p) \,\, , \label{eq:spectroscopic}
\eeq
in which $R(\omega_p)$ is the signal decay rate that depends only on the characteristic frequency of the manipulation sequence. 
By checking the $T$-dependence of the signal one can identify when the above approximation starts to work, and then one can use appropriate methods \cite{Alvarez_PRL11,Szankowski_PRA18} to obtain the values of $S(m\omega_p)$ for $m$ smaller than a certain finite $m_{0}$. Then, by changing $\omega_p$ one can perform a reconstruction of $S(\omega)$. 

The filter structure is analogous to the one obtained for periodic application of $\pi$ pulses in a Car-Purcell dynamical decoupling protocol \cite{Szankowski_JPCM17}, but we have now additional flexibility. The characteristic frequency $\omega_p$ is set by $\pi/(\tau+\delta t)$, while in the DD protocol it was equal to $\pi/\tau'$, with $\tau'$ being the interpulse time. We can then focus the filter at very low frequency by making $\delta t \! \gg \! \tau$, while not increasing the time that the qubit spends exposed to the noise. Formally, in the DD case the amplitude of $\delta$-like peaks was controlled by $|c^{\mathrm{DD}}_{m\omega_p}|^2 \! =\! \frac{4}{\pi^2 m^2}$, while in the considered protocol we have
\beq
|c_{m\omega_p}|^2 \approx \frac{\tau^2}{\delta t^2} \ll 1 \,\, ,
\eeq 
as long as $m \! \ll \! \delta t/\tau$, and $|c_{m\omega_p}|^2 \! \propto \! 1/m^2$ for $m \! \gg \! 2\delta t/\pi\tau$, which implies that in this large$-m$ regime $|c_{m\omega_p}|^2 \! \ll \!  \tau^2/\delta t^2$. 

When $\delta t \! \gg \! \tau$ one can thus tune the measurement-based frequency filter to very low $\omega_p$ by changing $\delta t$, while suppressing the coupling to this low-frequency noise by a factor controlled by $\tau/\delta t$, and these two tunings can be done practically independently. The most natural application of such a filter is when dealing with strong low-frequency noise that has some sharp spectral features, the characterization of which requires narrow filters, but one has to simultaneously suppress the amount of background noise picked up by filters of finite width. Let us discuss now such a situation at some length. 

%%%%%%%%%%%%%%%%%%%%%%%%%%%%%%%
%%% WHEN M CORRELATIONS BEAT DD
%%%%%%%%%%%%%%%%%%%%%%%%%%%%%%%
\subsection{Correlations of measurements vs dynamical decoupling: an example} \label{sec:example}
\begin{figure}[tb]
		\includegraphics[width=0.9\columnwidth]{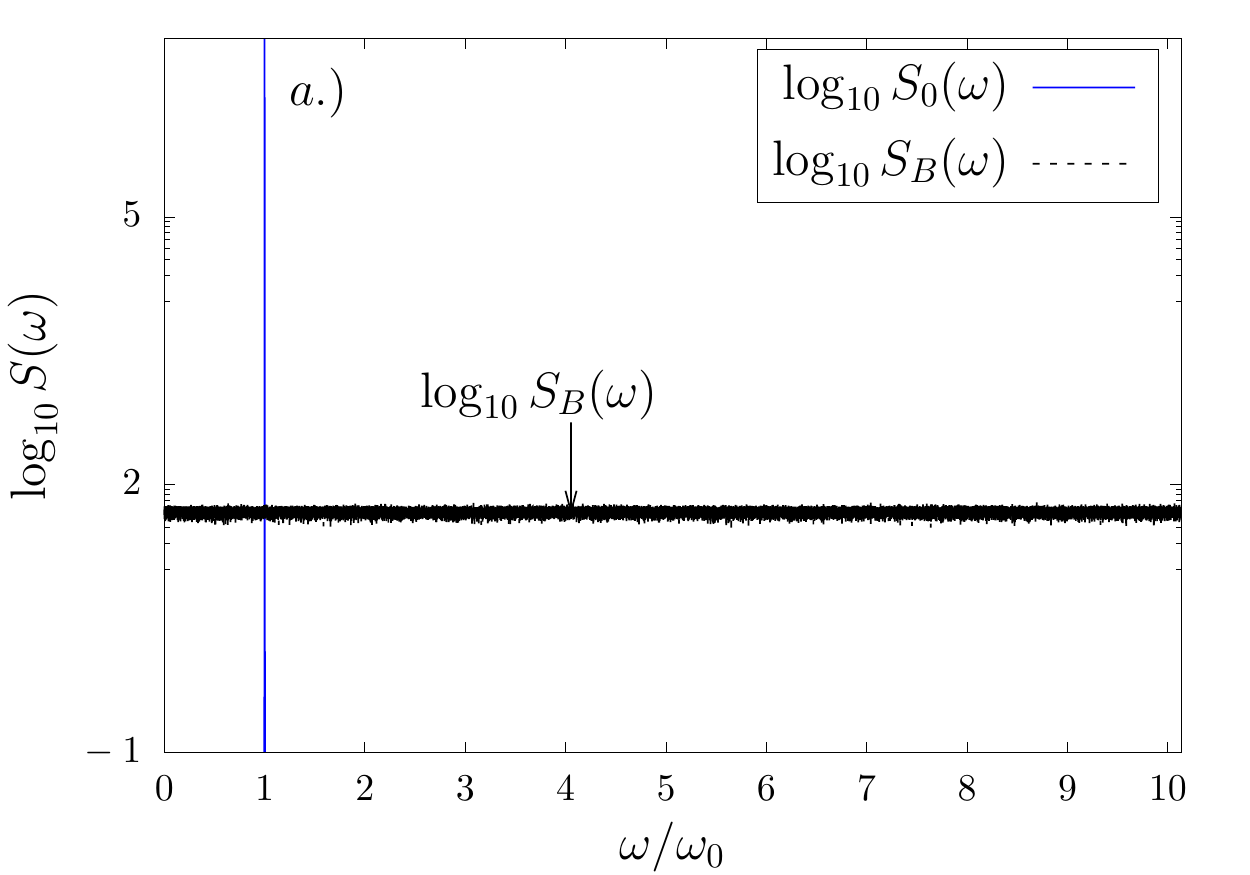}\hfill
		\includegraphics[width=0.9\columnwidth]{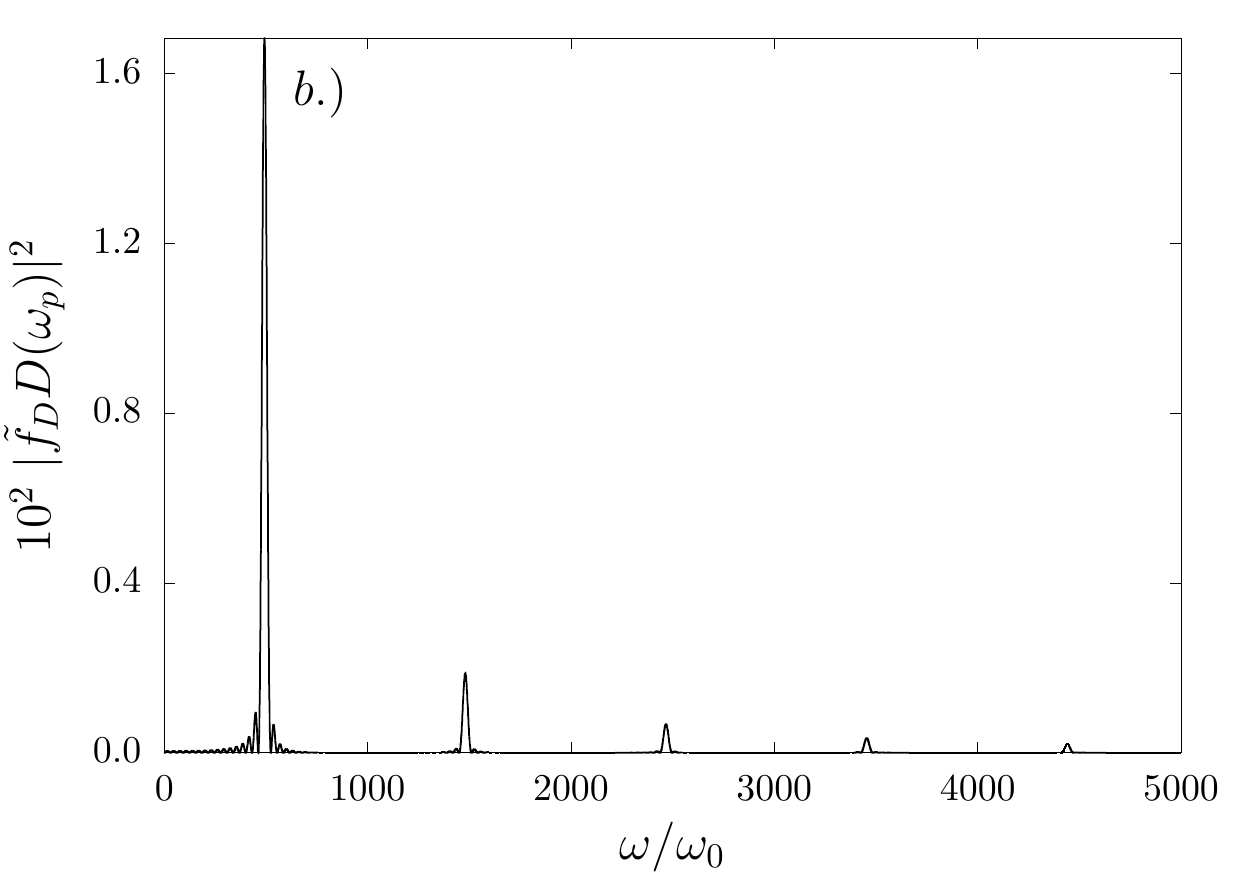}\hfill
		\includegraphics[width=0.9\columnwidth]{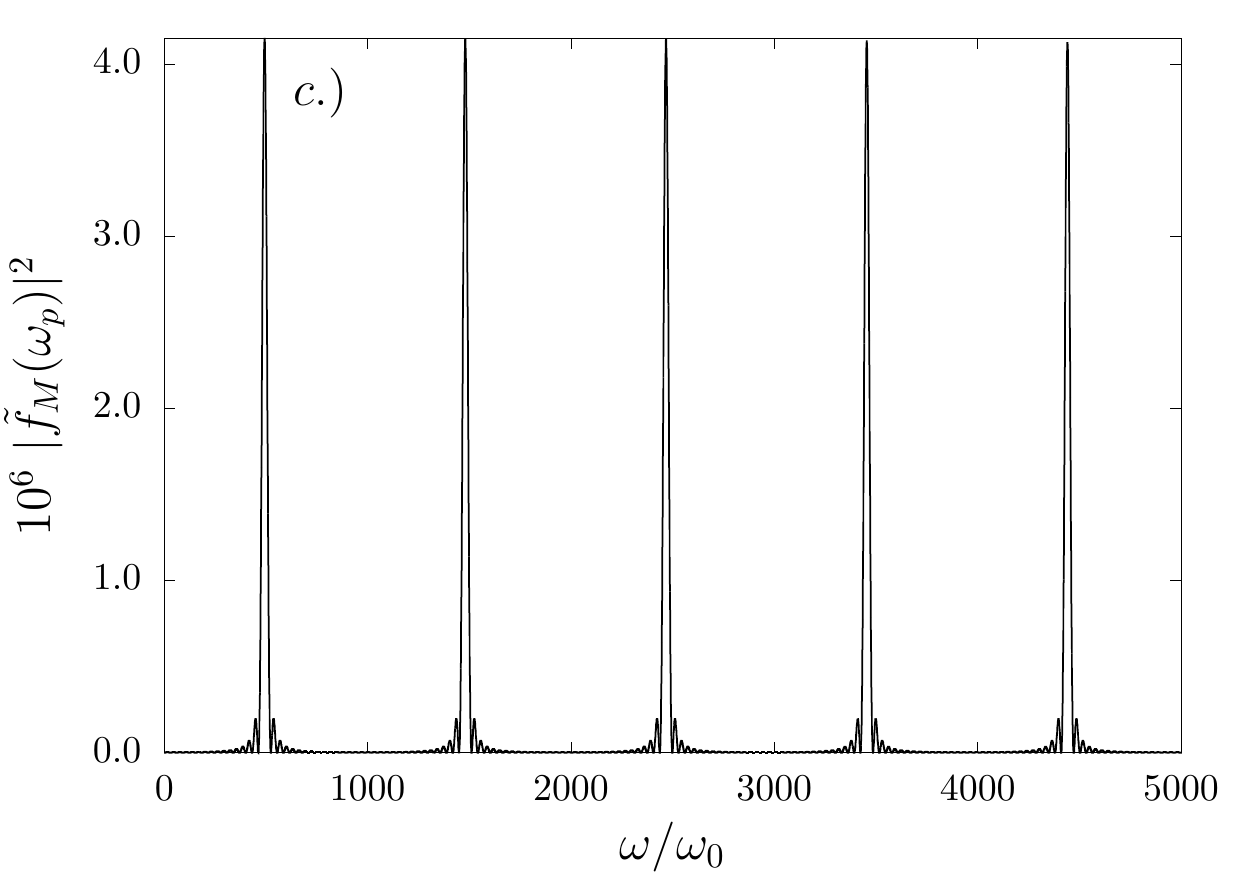}\hfill
	\caption{a) Model spectrum with a sharp peak at $\omega_0\approx 500$  (controlled by the height $\sigma^{2}_0=10^{7}$ and the width of the peak at $\omega\! = \! \omega_0,$ which is $\Delta\omega=1.5\times 10^{-2}$ with background noise $S_B\approx 50$ in the unit of the display). b) The DD filter with first peak matching $\omega_0$ c) Measurement filter for  $\tau/\delta t\approx 0.01 \ll 1$ with the first peak matching $\omega_0$.}\label{fig:spectrum_filters}
\end{figure}

We focus on an environment characterized by the power spectral density illustrated qualitatively in  Fig.~\ref{fig:spectrum_filters}a). The spectrum $S(\omega) \! =\! S_{0}(\omega) + S_{B}(\omega)$ consists of a very sharp spectral feature at frequency $\omega_0$, approximated by $S_{0}(\omega) \! \approx \! \sigma^{2}_{0} \delta(\omega-\omega_0)$ (in reality some sharply peaked function of width smaller than the width of filters that we are going to consider)  and a background $S_{B}(\omega)$ that is comparatively flat and featureless, i.e.,~$S_{B}(\omega) \approx S_{B} \! =$ const. 

Physical examples are most naturally found in the field of qubit-based characterization of small nuclear environments  (nanoscale nuclear magnetic resonance imaging) in which one tried to obtain precise information on precession frequency of one (or a few nuclei), in the presence of noise coming from many other nuclear spins \cite{Staudacher_Science13, Muller_NC14,DeVience_NN15,Lovchinsky_Science16,Boss_Science17,Schmitt_Science17}. Let us note that a filter that is equal to $0$ for most of the duration of the experimental protocol was
devised in a different experimental setting for exactly this purpose \cite{Laraoui_NC13} (see Sec. \ref{sec:Meriles} for a discussion).

Let us then assume that the main task of the spectroscopic procedure is to obtain the most accurate value of $\omega_0$. Consequently, we consider two spectroscopic procedures - one based on dynamical decoupling and the other on multiple measurements, characterized by the same total duration $T$, setting the characteristic frequency estimation precision to $1/T$. This means, that in the DD protocol the qubit is exposed to the environmental influence for time $T$, while in the measurement-based protocol it is exposed for time approximately equal to $T (\frac{\tau}{\delta t}) \! \ll \! T$, as we focus on the case of $\tau \! \ll \! \delta t$, for which we expect the qualitative difference between DD- and measurement-based filters to be most pronounced. 
We also assume that the characteristic base frequencies of both procedures, $\pi/\tau'$ and  $\pi/(\tau+\delta t)$, are both given by $\omega_p \! \approx \! \omega_0$. The frequency-domain filters corresponding to DD-based and measurement-based protocols are shown in Fig.~\ref{fig:spectrum_filters}b and Fig.~\ref{fig:spectrum_filters}c, respectively, for the case of $\omega_p \! =\! \omega_0$.

\begin{figure}[tb]
		\includegraphics[width=\columnwidth]{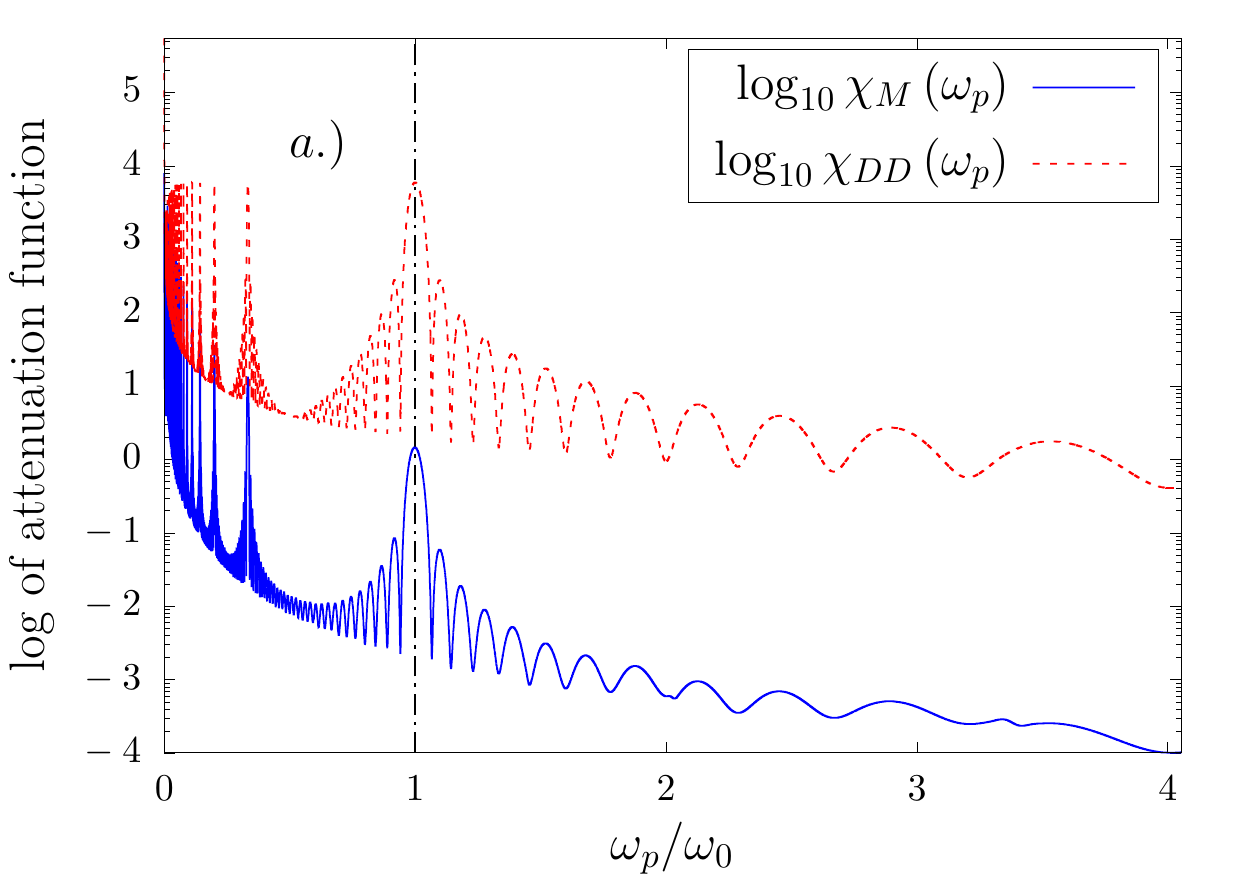}\hfill
		\includegraphics[width=\columnwidth]{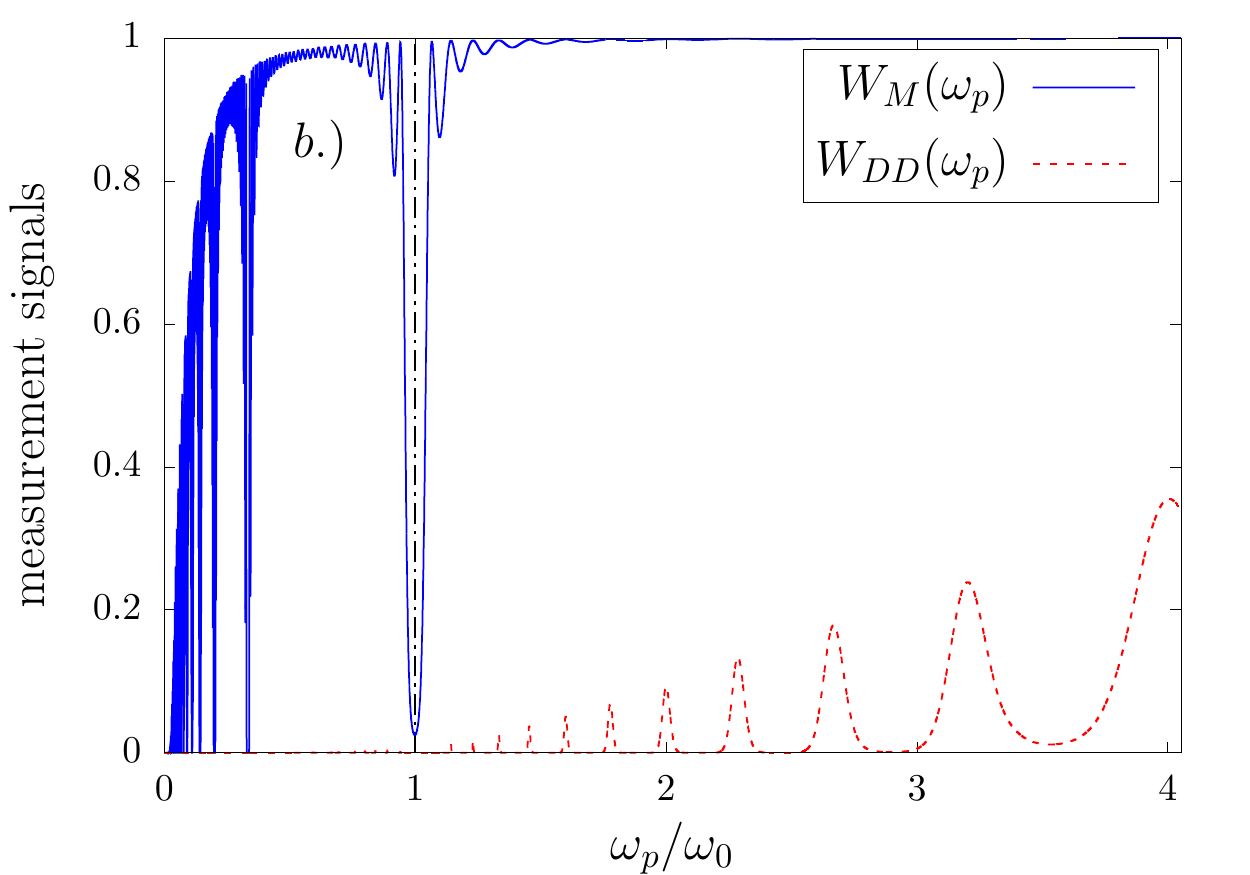}
	\caption{Comparison of a) attenuation functions and b) decoherence functions from DD and sequential measurement protocols for the model of the spectrum given in Fig. \ref{fig:spectrum_filters} with the same varying parameters $T_B=\frac{2\pi}{\omega_p}$, and $\tau=0.01T_B$, and a fixed numbers of interventions $N=16$. One can see a large peak in $\chi(\omega_p)$ at $\omega_p \! =\! \omega_0$, but note that all the odd sub-harmonics of $\omega_0$ frequency also give distinct features in the attenuation function. For the used set of parameters used, $\chi_{\mathrm{DD}}\! \gg \! \chi_{\mathrm{M}}$, and consequently the measured signal, $W\! \equiv \! \exp(-\chi)$, is basically zero in the DD case, while in the measurement-based protocol one can observe a clear collapse of the signal when $\omega_p$ is close to $\omega_0$.}\label{fig:chi_compare}
\end{figure}

For given $\omega_p$, the signal obtained from the DD-based (measurement-based) protocol is given by 
$\exp \left[ - TR^{D(M)}(\omega_p) \right]$ where the decay rate $R^{DD(M)}(\omega_p)$ is the sum of two contributions,
\beq
R^{\text{DD(M)}}(\omega_p) \equiv R^{\text{DD(M)}}_{0}(\omega_p) + R^{\text{DD(M)}}_{B}(\omega_p) \,\, ,
\eeq
the first coming from the sharp spectral feature centered at $\omega_0,$ being non-zero for $|\omega_p - \omega_0| \lesssim \frac{1}{T}$, and the second from the background noise spectrum. The precise estimation of $\omega_0$ by tuning $\omega_p$ is possible, when $\chi_{B} \! \ll \! 1$ in the scanned frequency range, while $\chi_{0}$ changes from much less than $1$ to a value $\approx 1$ when $\omega_p \! = \! \omega_0$. When the latter maximal value of $\chi_0$ is much greater than $1$ the estimation of the value of $\omega_0$ is still possible, but the precision will be lower, as the measured signal could become unmeasurably small in the whole range of $|\omega_p - \omega_0| \lesssim \frac{1}{T}$.

The contribution of the background spectrum to the signal decay rate is given in the case of the dynamical decoupling protocol by
\beq
R^{\text{DD}}_{B} \approx \frac{4}{\pi^2}S_{B} \,\, ,
\eeq
where we have taken into account only the first peak of the filter (as $|c_{m\omega_p}|^2 \propto 1/m^2$ and the background spectrum is qualitatively flat), while in the measurement-based scheme we have
\beq
R^{\text{M}}_{B} \approx   \sum_{m} |c_{m\omega_p}|^2  S_{B}(m\omega_p) \approx  \frac{\tau}{\delta t}S_B  \ll R^{\text{DD}}_{B} 
\eeq
where we have taken into account that up to $m\approx \delta t/\tau$ we have $|c_{m\omega_p}|^2 \! \approx \! (\tau/\delta t)^2$ while for larger $m$ these coefficient become much smaller, and than $S_{B}(\omega)$ is assumed to be flat (or at least not strongly increasing) up to $\omega \! \approx \omega_p \times \delta t/\tau$. 

The background contribution to the measurement-based signal is thus negligible when $S_{B}T\tau/\delta t \! \ll \! 1$, while it completely dominates the DD-based signal when $S_{B}T \! \gg \! 1$. The qualitative difference between its effect on the two protocols is present when $\delta t/\tau \! \gg \! S_{B}T \! \gg \! 1$. 

We see that for an approximately flat $S_B(\omega)$, the background contribution to the signal is much smaller in the measurement-based protocol, compared to the DD-based one. This is due to the diminished noise sensitivity of the latter. It remains to be checked, whether this diminished sensitivity does hinders our observation of the signal related to the sharp spectral feature $S_0 (\omega)$.
Assuming that the width of the filter peak centered at $\omega_p$ is still larger than the width of this feature, the maximal contribution to the attenuation function from the $S_{0}(\omega)$ part of the total spectrum is $\chi^{\max}_{0} \! \approx \! \sigma^2_{0} T^2 |c_{\omega_p}|^2$ when the filter peak is centered at $\omega_0$ \cite{Szankowski2019}.  This contribution is approximately equal to $1$ in the measurement-based protocol when $\delta t / \tau \approx \sigma_0 T$. Taking into account the previously derived inequality, we see that the measured-based protocol can significantly outperform the DD-based one in locating $\omega_0$ with $1/T$ accuracy if $\sigma_{0} \! \gg \! S_{B}$, i.e.,~when the sharp spectral feature strongly dominated over the background spectrum in the vicinity of $\omega_0$, and if the desired frequency resolution is $1/T \! \ll \! S_{B}$. In Fig.~\ref{fig:chi_compare} we illustrate this with calculations of $\chi(\omega_0)$ and the corresponding observables $W\equiv \exp(-\chi)$ for the two protocols.  

It might seem that the width of the filter peaks corresponding to the measurement sequence could be made arbitrarily small, resulting in arbitrarily high precision of the estimation of $\omega_0$ \cite{Boss_Science17,Schmitt_Science17}, as making $\delta t$ longer seems not to have any downside: We just have to wait longer before we reinitialize the qubit. However, $\delta t$ cannot be in reality be made arbitrarily long; in order for the above formulas to describe the experimental results involving many repetitions of sequences of measurements, the timing of measurements and reinitializations has to be very stable over a timescale much longer than that of a single $\delta t $ delay \cite{Boss_Science17,Schmitt_Science17,Gefen_PRA18}. If $\delta t$ fluctuates by an amount $\sigma_{\delta t}$ amount over ``macroscopic'' timescale of data acquisition in the experiment, the positions of filter peaks fluctuate by this amount, and their width after averaging becomes approximately equal to $\sigma_{\delta t}$, not $1/T$. The precision of the $\omega_0$ frequency estimation in the measurement-based protocol is thus limited by the stability of the classical clock, according to which all the operations in the whole experiment are conducted \cite{Boss_Science17,Schmitt_Science17,Gefen_PRA18}.

%%%%%%%%%%%%%%%%%
% OTHER PROTOCOLS
%%%%%%%%%%%%%%%%%
\section{Related single-qubit protocols}  \label{sec:related}
We have shown how, by considering correlations of measurements of $\hat{\sigma}_x$ and $\hat{\sigma}_y$ one can modify the influence of environmental noise $\xi(t)$ by effectively multiplying it by a time-domain filter $f(t)$ that is equal to zero during $\delta t_k$ periods of time. Let us now discuss earlier works, in which other projective measurements were considered \cite{Bechtold_PRL16}, or a completely different (not measurement-based) method of creation of such an $f(t)$ was used \cite{Laraoui_NC13}. We will also discuss how the previously obtained results apply to an experiment in which the qubit is not re-initialized in a fixed state after each measurement.

%%%%%%%%%%%%
%%% BECHTOLD
%%%%%%%%%%%%
\subsection{Projections only on $\ket{+x}$} \label{sec:Bechtold}
In \cite{Bechtold_PRL16} correlations of measurements on a quantum-dot-based spin qubit were measured and discussed. Contrary to the measurements discussed here so far, those were ideal negative measurements \cite{Emary2014}: A spin-selective optical excitation was applied on the qubit at $t_k$ times, the result being either destruction of the qubit for one spin direction or leaving it in the opposite spin state (denoted by $\ket{\downarrow}$ in \cite{Bechtold_PRL16} and as $\ket{+x}$ here, in order to maintain a closer connection with the rest of the paper). We can thus identify the measurement used there with projection on the $\ket{+x}$ state. The probability of successfully performing of such a projection after the $k$-th period of evolution of the qubit is given by $p_{x}(+\vert \alpha_k)$ from Eq.~\eqref{eq:prob_single_x}. The spin qubit was subjected in \cite{Bechtold_PRL16} to up to three such projections at times $t_0\! = \!0$, $t_1$, and $t_2+t_1$, with  the first projection at $t_0 \! =\! 0$ used to  initialize the qubit. With $\langle\cdots\rangle$ denoting averaging over noise with the qubit initialized in the $\ket{+x}$ state, and that $\hat{P}_{+x} = \frac{1}{2}(\openone+\hat{\sigma}_x)$, the expectation value of the projection at $t_1$ (or a correlation between projections at $t_0 \! =\! 0$ and $t_{1}$ in the context of the experiment \cite{Bechtold_PRL16}) is given by
\begin{align}
C_{+x}(t_1) & :=  \frac{1}{2}\langle(\openone+\hat{\sigma}_x(t_1))\rangle =  \frac{1}{2}(1+\langle\cos\alpha_1\rangle) \nonumber\\
& = \frac{1}{2}[1 + C_{x}(t_1,t_1)]\,\, .
\end{align}
in which $C_{x}(t_1,t_1)$, defined in Eq.~\eqref{eq:Cxxxx}, is the expectation values of $\hat{\sigma}_x$ discussed in the rest of the paper, while the quantity $C_{+x}(t_1)$ defined above  was called $g_2(t_1)$ in \cite{Bechtold_PRL16}. The correlation of projections at $t_1$ and $t_2$, called $g_{3}(t_1,t_{2})$ in \cite{Bechtold_PRL16}, is 
\begin{align}
C_{+x,+x}(t_1,t_2) & := \frac{1}{4}\langle[\openone+\hat{\sigma}_x(t_1))(\openone+\hat{\sigma}_x(t_2)] \rangle  \,\, , \nonumber\\
& \!\!\!\!\!\!\!\!\!\!\!\!\!\!\!\!\!\!\!\! = \frac{1}{4}[1+ C_{x}(t_1,t_1) + C_{x}(t_2,t_2-t_1) + \langle\cos\alpha_1\cos\alpha_2\rangle ]\,\, . 
\end{align}
Using now the fact that $\cos\alpha_1\cos\alpha_2 \! =\! \frac{1}{2}\cos(\alpha_1+\alpha_2) + \frac{1}{2}\cos(\alpha_1-\alpha_2)$ we obtain
\begin{align}
C_{+x,+x}(t_1,t_2) & = \frac{1}{4}[1+ C_{x}(t_1,t_1) + C_{x}(t_2,t_2-t_1) \nonumber\\
& + \frac{1}{2}\mathrm{Re}g(1,1) + \frac{1}{2}\mathrm{Re}g(1,-1) ] \,\, , \label{eq:Cpp}
\end{align}
where we recognize the echo signal $g(1,-1)$ given previously in Eq.~(\ref{eq:echo}). It should become clear now that the measurement setup used in \cite{Bechtold_PRL16} gives results qualitatively the same as the setup considered here: The correlation of $n$ projections on $\ket{+x}$ is related to a linear combination of coherence signals obtained with $k\! < \! n$ pulses. The observable feature of observable from \cite{Bechtold_PRL16} that distinguishes it from the correlation functions considered in this paper is the appearance of signals corresponding to the evolution for a fraction of the total protocol time, e.g.,~$ C_{x}(t_2,t_2-t_1)$ in Eq.~(\ref{eq:Cpp}).

%\begin{align}
%C_{+x,+x}(t_1,t_2) & = \frac{1}{2}[ C_{+x}(t_1)+C_{+x}(t_2)+\frac{1}{2}C_{+x}(t_1+t_2)] -\frac{3}{8} \nonumber\\ 
%& + \frac{1}{8}g(1,-1) \,\, ,  \label{eq:Cpp}
%\end{align}

%%%%%%%%%
% MERILES
%%%%%%%%%
\subsection{Sequences consisting of both $\pi$ and $\pi/2$ pulses} \label{sec:Meriles}
		\begin{figure}[tb]
			\centering
			\includegraphics[scale=1]{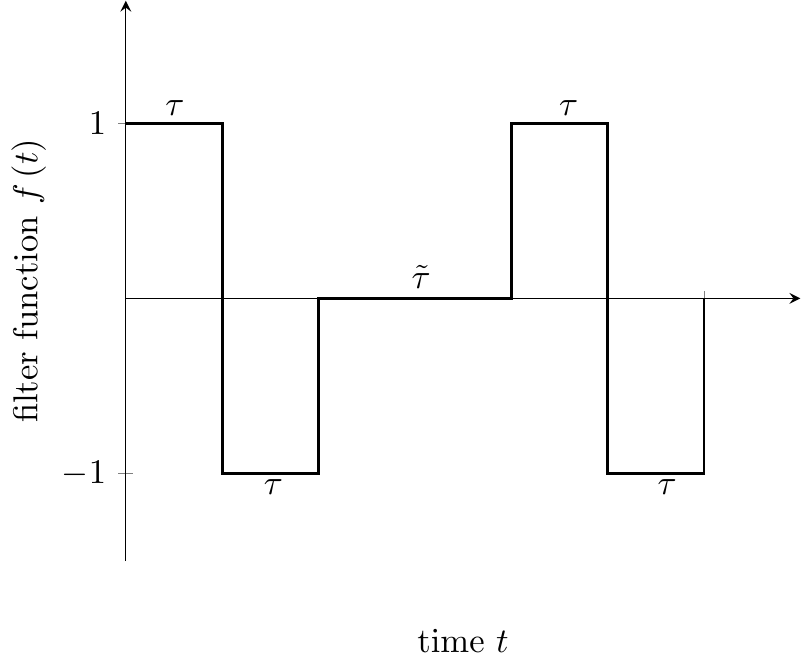}
			\caption{Filter function implemented in \cite{Laraoui_NC13} with the use of both $\pi$ and $\pi/2$ pulses acting on the qubit (in the notation used in that work, $\delta t \! = \! \tilde{\tau}$).}\label{fig:f_Mer}
		\end{figure}
		
A purely pulse-based way to obtain a temporal filter $f(t)$ that is equal to zero for an adjustable period of time was described in \cite{Laraoui_NC13}. The correlation spectroscopy protocol considered there was the following: The qubit was initialized in the $\ket{+x}$ state, subjected to a $\pi$ pulse after time delay $\tau$, and then at time $2\tau$ (at the echo rephasing time) it was subjected to a $\pi/2$ pulse that was converting the relative phase between $\ket{\pm z}$ states into the occupation $p_{+}$ of the $\ket{+z}$ state (with the occupation of the other state, $\ket{-z}$, given by $p_{-} \! =\! 1-p_{+}$). These occupations were then immune to the dephasing noise acting along the $z$ axis on the qubit - they could be perturbed only by much slower processes of energy exchange between the qubit and the environment (spin-phonon scattering in the case of nitroge-vacancy center used in \cite{Laraoui_NC13}). The state of the qubit could then be considered frozen for time $\delta t$ (called $\tilde{\tau}$ in \cite{Laraoui_NC13}), if only $\delta t$ was much shorter than the qubit energy relaxation time $T_{1}$. After this delay $\delta t$, the qubit was rotated again onto the equator of the Bloch sphere, and subjected to the second echo sequence, characterized by delay $\tau$. The filter function corresponding to such an experiment is given then by the filter function given in Fig.~\ref{fig:f_Mer}. Clearly, using techniques described in \cite{Laraoui_NC13} one can construct filters that consist of DD parts (oscillating between $1$ and $-1$) separated by periods $\delta t$ during which the filter is zero. However, as already mentioned, $\delta t \! \ll \! T_1$ is required, while in the measurement-based protocols described in this paper, $\delta t$ is limited only by the timescale on which the classical clock loses its stability \cite{Boss_Science17,Schmitt_Science17,Gefen_PRA18} (see Sec. \ref{sec:example}).

%%%%%%%%%%%%%%%%%%%%%%%% 
%%% NO RE-INITIALIZATION
%%%%%%%%%%%%%%%%%%%%%%%%
\subsection{Protocol without re-preparation of qubit state after measurement} \label{sec:noreinit} 
So far we have considered protocols in which the qubit is found in a known state at the beginning of each $\tau_k$ evolution period, either due to reinitialization, or due to the fact that the measurement succeeds only when the qubit is found in a given state, as in Sec.~\ref{sec:Bechtold}. Let us now show that such a repreparation in a given fixed state is not necessary to connect the expectation value of the measurement on the qubit with noise filtering. 

For simplicity, let us consider only the case of only $\hat{\sigma}_x$  measurements. If at the beginning of the $k$-th period of interaction with the environment the qubit is in state $r\! =\! \pm$ (one of eigenstates of $\hat{\sigma}_{x}$), the probability of obtaining the result $m$ for measurement after time $\tau_k$ is 
		\begin{align*}
			p_x\cv{m\vert\alpha_k,r}=\dfrac{1}{2}\cv{1+m\cdot r\cos\alpha_k} \,\, ,%\\
			%p_y\cv{m\vert\alpha_k,p}=\dfrac{1}{2}\cv{1+m\cdot r\sin\alpha_k}
		\end{align*}
		which is a straightforward generalization of Eq.~\eqref{eq:prob_single_x}.
	In this more general setting it is clear that the measurement gives information on the difference in sign between the measured and the initial state. In other words, we can write $p_x\cv{m\vert\alpha_k,r}=p_x\cv{mr\vert\alpha_k,+}.$

For sequences of outcomes $\cv{m_1,\ldots,m_n}$ and repreparations $\cv{r_1,\ldots,r_n},$ the probability of measuring the former is then
			\begin{equation}
					p\cv{m_1,\ldots,m_n\vert r_1,\ldots,r_n} = \prod_{k=1}^n p_{x}\cv{m_k\vert\alpha_k,r_k} \,\, ,
			\end{equation}
		which will coincide with Eq.~\eqref{eq:p_reinit} if all prepared states are $r_k\!= \!+$. Let us consider now the protocol without re-preparation, in which the qubit state at the beginning of the $k$-th period of interaction with the environmental noise is precisely the state obtained by projective measurement at the end of the previous period,~i.e., $r_1=+$ and $r_k\! =\! m_{k-1}$ for $1\! <\! k \leq n$. We consider now the expectation value of only the last measurement, all the previous measurements are performed, but their results are discarded and averaged over. Denoting $p_{x}(m_k\vert \alpha_k,+)$ by $p_{k}(m_k),$ we have 
\begin{align}
&\hspace{-0.5cm}\aveb{m_n}_{\text{no re-prep}}\nn\\ & = \sum_{m_n,\ldots,m_1}m_{n} p(m_1,\ldots,m_n \vert _,m_1,\ldots m_{n-1}) \nonumber\\
& = \sum_{m_n,\ldots,m_1} m_n p_1(m_1) p_2(m_2 m_1) \ldots p_n(m_n m_{n-1})  \nonumber\\
& = \sum_{m_n,\ldots,m_1} (m_n m_{n-1})(m_{n-1}m_{n-2}) \ldots (m_2 m_1) m_1 \nonumber\\
& \hspace{1.5cm}\times p_1(m_1)p_2(m_2m_1)  \ldots p_n(m_n m_{n-1}) \nonumber\\
& = \sum_{m_n',\ldots,m_1'} \prod_{k=1}^{n} m_{k}' p_k(m_k') \,\, \label{eq:mn}
\end{align}		
where in the last expression, using the fact that $m_k^2 \! =\! 1$, we recognize the correlation $\langle m_1,\ldots, m_n\rangle$ in the case of repreparation of the qubit state in $\vert+x\rangle$, which is the quantity discussed throughout this paper.

%CONNECT TO OPERATIONS PAPER NOW...
The result that an expectation value of the last measurement in a sequence without state repreparation is equivalent to a correlation of all measurements in a sequence with repreparation of qubit state can in fact be derived under more general assumptions. In \cite{Sakuldee_operations} we show that it holds also for the environments described quantum mechanically, while here we have focused on the case of environment as a source of classical noise. However, while other general results from \cite{Sakuldee_operations} on relations between the effects of a sequence of measurements on the qubit, and a sequence of unitary dynamical decoupling operations (the relation between correlations of measurements and noise filtering, on which we focus here, being one example of such a relation) hold for a completely general form of qubit-environment coupling, the connection between protocols with and without repreparations holds only for the pure dephasing case.

%%%%%%%%%%%%%%%
%%% NONGAUSSIAN
%%%%%%%%%%%%%%%
\section{Witnessing non-Gaussian character of noise} \label{sec:nonGaussian}
We start with a following observation: For Gaussian noise we obtain $C_{xy}$ given by
	\begin{equation}
	\begin{split}
			C_{xy}&(t_1,\tau_1; t_2, \tau_2)\\ &= \frac{1}{2} \left[ \sin\Omega(\tau_1 + \tau_2) e^{-\chi_{1,1}} +\sin\Omega(\tau_2 -\tau_1) e^{-\chi_{1,-1}} \right] \,\, , \label{eq:Cxy}
	\end{split}
	\end{equation}
	which is equal to zero in the rotating frame, in which we can set $\Omega \! =\! 0$.
If we do not make any assumption about the statistics of the noise, we have a general expression for the $x-y$ correlation in the rotating frame
	\begin{align}
		C_{xy}(t_1,\tau_1; t_2, \tau_2)		&= \dfrac{1}{4i}\left(\aveb{e^{i\cv{\Phi_2+\Phi_1}}}-\aveb{e^{-i\cv{\Phi_2+\Phi_1}}}\right.\nonumber\\
		&\left.\hspace{1cm}+\aveb{e^{i\cv{\Phi_2-\Phi_1}}}-\aveb{e^{-i\cv{\Phi_2-\Phi_1}}}\right) \,\, ,\label{eq:Cxy_gen}
	\end{align}
	which is zero not only in the case of Gaussian noise, but also for all the non-Gaussian noises with vanishing odd cumulants. In the latter case the averages $\aveb{e^{i\Phi_{1,s_2,\ldots,s_n}}}$ and $\aveb{e^{-i\Phi_{1,s_2,\ldots,s_n}}}$ will coincide with $\exp\cv{\displaystyle\sum_{k=1}^\infty\cv{-1}^k\chi^{2k}_{1,s_2,\ldots,s_n}},$ where $\chi^{k}_{1,s_2,\ldots,s_n}$ is related to the $k\text{th}$ cumulant $K\cv{\xi\cv{t_1}\cdots\xi\cv{t_k}}$ of the noise \cite{Szankowski_JPCM17,Norris_PRL16} via
	\begin{equation}
		\begin{split}
		\chi^{k}_{1,s_2,\ldots,s_n} &= \dfrac{1}{k!}\int_0^{t_n}d^kt_1\cdots\int_0^{t_n}d^kt_k\\
		&\hspace{1cm}\times\cv{\prod_{i=1}^kf_{1,s_2,\ldots,s_n}\cv{t_i}}K\cv{\xi\cv{t_1}\cdots\xi\cv{t_k}}.\label{eq:cumulant}
		\end{split}
	\end{equation}

\begin{figure}[tb]
	\begin{center}
		\includegraphics[width=\linewidth]{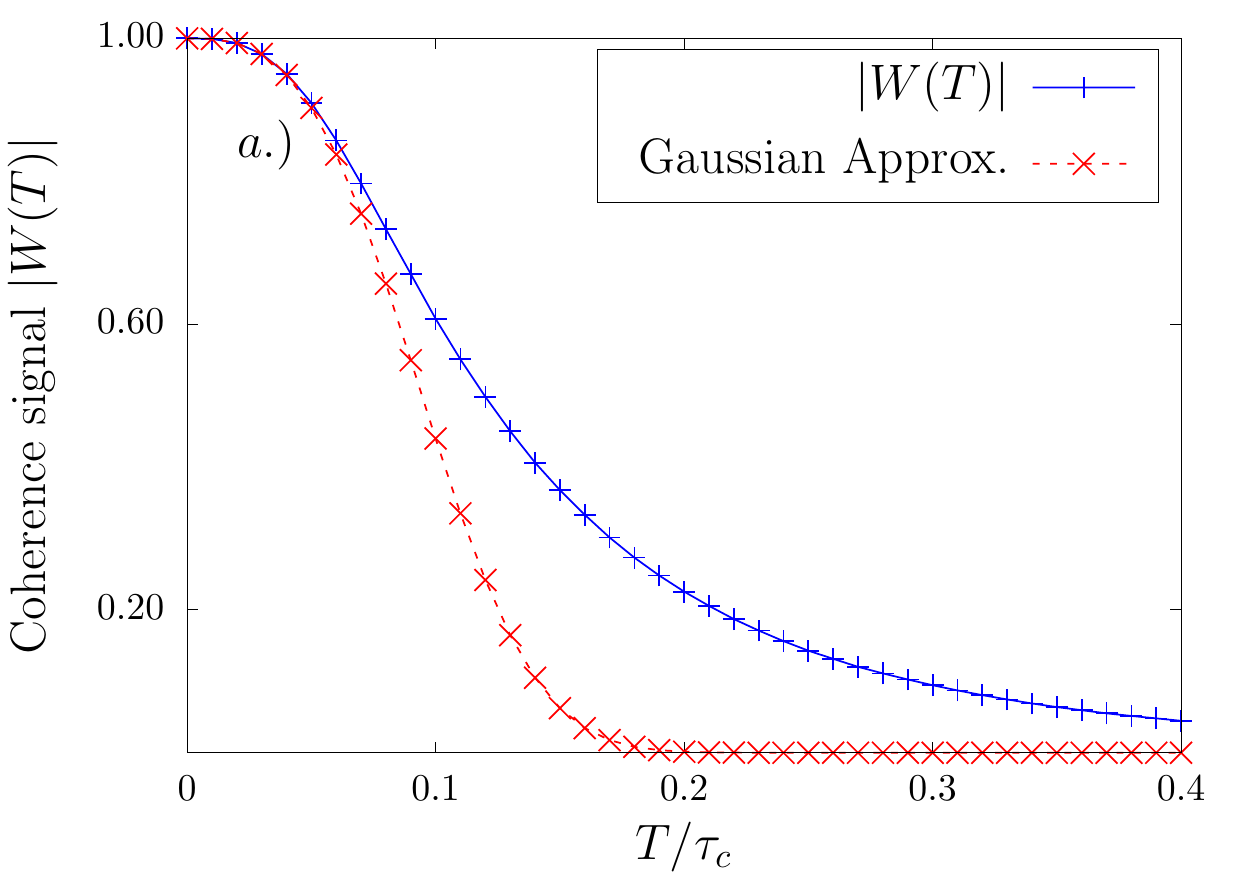}
	\end{center}
	\hfill
	\begin{center}
		\includegraphics[width=\linewidth]{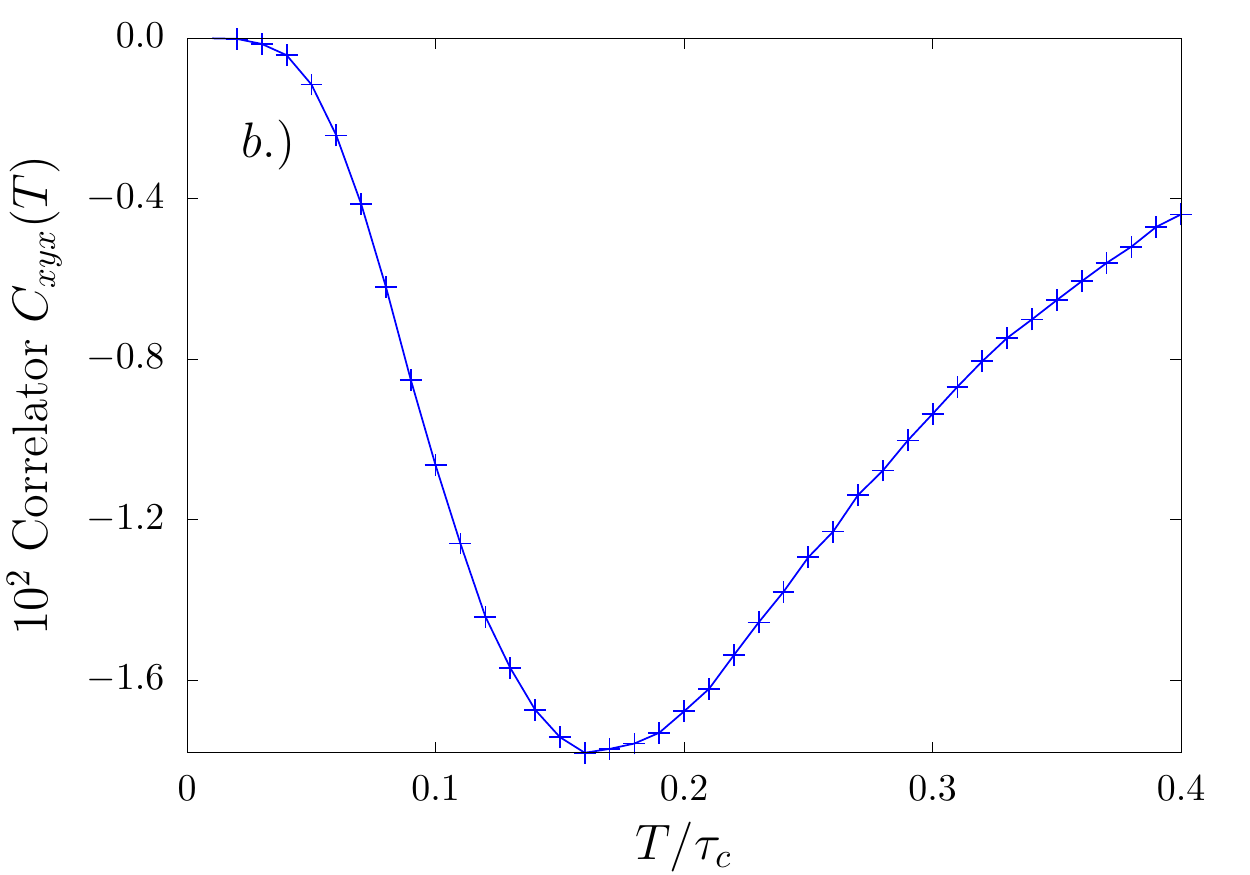}
	\end{center}
	\caption{Coherence calculation for quadratic coupling to Ornstein-Uhlenbeck noise with $v_2\! =\! 100/\tau_c$. Time is in unit of $\tau_c$. a) $|W|$ for the CP-2 sequence. The red dashed line is the Gaussian approximation; the total signal is clearly non-Gaussian b) Imaginary part of $W$ for CP2. }\label{fig:ImCP2}
\end{figure}
	
Even though the vanishing of $C_{xy}$ does not necessarily imply the  Gaussian statistics of the noise, it can be used as evidence of non-Gaussianity. In other words, if the correlator $C_{xy}\neq 0$, it means that the environmental noise is non-Gaussian.
	
Moreover, instead of focusing  on the second -order correlator $C_{xy}$, one can also look at correlators involving a larger number of measurements, with an odd number of them along the $y$ axis. A simple and particularly useful correlator in the following example is the $C_{xyx} \! =\! \langle \cos\alpha_1 \sin\alpha_2 \cos\alpha_3 \rangle$ with $\tau_1 \! =\! \tau_3 \! = \!\tau$ and $\tau_2 \! = \! 2\tau$, i.e.,~the timing pattern corresponding to the two-pulse Carr-Purcell sequence when all $\delta t_{k}\! =\! 0$.   
In the presence of strong low-frequency noise, and for qubit-noise coupling time exceeding the $T_{2}^{*}$ time of dephasing under free evolution, $\tau \gg \! T_{2}^{*}$, the only contributions to $C_{xyx}$ that correspond to balanced filters give a non-vanishing signal and we have
\begin{equation}
C_{xyx} \approx -\frac{1}{4}\mathrm{Im}\left\langle e^{i(\Phi_{1}-\Phi_2+\Phi_3)} \right\rangle \,\, ,    \label{eq:Cxyx}
\end{equation}
which is simply equal to the imaginary part of the coherence signal of the qubit subjected to the CP-2 sequence.

As an example let us consider now an often-encountered situation in which the environmental noise is non-Gaussian, that of quadratic coupling to Gaussian noise, i.e.~when the coupling of the qubit to noise is given by $v_{2}\hat{\sigma}_{z}\xi^{2}(t)$. This occurs when the qubit is at the so-called optimal working point (or at the clock transition, using the terminology of atomic physics), at which the first derivative of its energy splitting $\Omega$ with respect to the dominant noisy parameter is zero and the second-order term in the Taylor expansion of $\Omega$ has to be taken into account. Such a working point was found for many types of qubits \cite{Ithier_PRB05,Yoshihara_PRL06,Petersson_PRL10,Wolfowicz_NN13,Medford_PRL13,Malinowski_PRB17}. Another situation in which quadratic coupling to noise appears is when we have a transverse coupling to noise, $v_{x} \hat{\sigma}_{x}\xi_{x}(t)/2,$ in the presence of large static splitting $\Omega \hat{\sigma}_{z}/2$ [and possibly some longitudinal noise $\hat{\sigma}_{z}\xi_{z}(t)/2$]. When $\langle \xi^2_{x}\rangle^{1/2} \! \ll \! \Omega$, the qubit evolution is effectively of pure dephasing character, with an additional contribution to noise along the $z$ axis given by $v_{2}\hat{\sigma}_{z} \xi^{2}_{x}(t)$ with $v_{2} \! = v^2_x/4\Omega$ \cite{Cywinski_PRB09,Szankowski_QIP15}. 

While $\xi(t)$ is assumed now to be Gaussian, $\xi^2(t)$ is a non-Gaussian process \cite{Makhlin_PRL04,Cywinski_PRA14} that has nonzero odd cumulants when filtered by a DD sequence consisting of an even number of pulses \cite{Cywinski_PRA14}. Consequently, the imaginary part of the CP-2 coherence signal, and the above $C_{xyx}$, is nonzero due to the non-Gaussian character of such a noise. As an example, we consider the case of $\xi_x(t)$ being an Ornstein-Uhlenbeck process with correlation time $\tau_c$, rms $\sigma \! =\! 1$, and coupling to the qubit $v_2 \! =\! 100/\tau_c$. In Fig.~\ref{fig:ImCP2}a) we show the decay of the CP-2 coherence signal as a function of total sequence time $T\! \equiv \! 4\tau$ obtained from numerical simulation of the qubit subjected to many realizations of such noise. We show there also the Gaussian result, in which only the second cumulant of noise is included in the calculation of $\chi(T)$. In Fig.~\ref{fig:ImCP2}b) we show the result for $C_{xyx}$ from Eq.~(\ref{eq:Cxyx}). This will correspond to the measured signal if the qubit is additionally exposed to very low-frequency longitudinal noise leading to $T_{2}^{*}\! \ll \! 0.1 \tau_{c}$. This is the situation encountered, for example, for spin qubits in quantum dots, which are exposed to both longitudinal and transverse noise due to nuclear spins, and the longitudinal noise is concentrated at much lower frequencies than the transverse one \cite{Malinowski_PRL17}.

%%%%%%%%%%%%%%
%%% CONCLUSION
%%%%%%%%%%%%%%
\section{Discussion and Conclusion}  \label{sec:conclusion}
We have investigated here the protocol, in which a qubit, experiencing pure dephasing due to an environment that can be treated as a source of external classical noise, is subjected to multiple initializations, evolutions, and projective measurements. The expectation values of correlations of multiple measurements can be expressed in a form closely related to the one that describes the dephasing of the qubit subjected to a dynamical decoupling sequence of pulses. More precisely, for $n$ measurements, each along the $x$ or $y$ axis of the qubit, from a linear combination of approximately$2^{n}$ correlators one can construct an observable that is equal to the dynamical decoupling signal obtained by performing a $\pi$ rotation of the qubit at each of the measurement times. Conversely, a single correlator of $n$ measurements is a linear combination of $2^{n}$ dynamical decoupling signals, corresponding to sequences in which at each measurement time a $\pi$ pulse is either applied or not. This generalizes the result of \cite{Fink_PRL13}, where it was noticed that correlation of two measurements of $\hat{\sigma}_{x}$ of the qubit corresponds to a linear combination of free evolution and echo signals. 

The above relationship between correlators of multiple measurements and dynamical decoupling signals offers the possibility of performing noise spectroscopy \cite{Degen_RMP17,Szankowski_JPCM17} of Gaussian and non-Gaussian noises without applying any $\pi$ pulses to the qubit, only by repeatedly initializing it and measuring. It should be noted that repetitive measurements on the qubit interacting with an environment, done in the quantum Zeno regime in which the environmental perturbation of the initial state is small, also lead to observables determined by expressions involving filtered environmental noise \cite{Zwick_PRAPL16,Do2019,Muller_arXiv19}, or give more qualitative insight into autocorrelation time of the environmental dynamics \cite{Muller_SR16}. The protocols discussed here do not however rely on the small-perturbation regime, and they also involve multiple re-initializations of the qubit (note however that we have also discussed a version of the protocol without such re-initializations).
The family of frequency filters obtained in this way is also richer than the one that is relevant for dynamical decoupling - the possibility of having long periods of time in which the time-domain filter is zero allows for more flexibility in focusing the filter at low frequency features of noise, without leading to complete suppression of the observable signal due to exposure to other frequencies. Note that such filters were discussed previously, but using a different control protocol, in which both $\pi$ and $\pi/2$ pulses were used \cite{Laraoui_NC13}. However, in that protocol the time $\delta t$ of the qubit being insensitive to the environmental noise was limited by the time $T_{1}$ time of the qubit, while in the measurement-only scheme the only limitation is the time after which the classical clock used to time all the operations in the protocol loses its stability \cite{Boss_Science17,Schmitt_Science17,Gefen_PRA18}. Finally, we have shown how a correlation of three measurements can be used as evidence of a non-Gaussian statistics of the environmental noise.

We stress that our focus here was solely on the case in which the environment can be treated as a source of classical noise, the stochastic dynamics of which is independent of the existence of the qubit. The effects of backaction of the measurement on the qubit on the state of a mesoscopic \cite{Fink_arXiv14,Bethke_arXiv19} or 
small quantum environment, e.g.,~a single nuclear spin, coupled to it have been a subject of intense recent attention \cite{Ma_PRA18,Gefen_PRA18,Pfender_NC19,Cujia_Nature19}, but these are beyond the scope of this paper. 
Let us however discuss briefly some open problems in the ongoing research on qubit-based noise spectroscopy (more generally, qubit-based environment characterization, and the elucidation of them by this paper and a related work \cite{Sakuldee_operations}. An arguably key foundational problem in this field is gaining a better understanding of conditions, under which one can treat the environmental noise as effectively classical stochastic signal affecting the qubit. Quantum features of Gaussian noise can be sensed with multiple qubits or qubits that have only one of their energy levels coupled to the environment \cite{Paz_PRA17,Kwiatkowski_phase_arXiv19}. The non-Gaussian nature of environmental noise can also be witnessed with methods other than the one discussed here \cite{Zhao_PRL11,Norris_PRL16,Kwiatkowski_PRB18,Sung2019,Ramon_PRB19}. However, discerning between effectively classical and truly quantum noise affecting the qubit that undergoes pure dephasing in a general non-Gaussian case remains elusive, although progress is being made in this direction \cite{Reinhard_PRL12,Hernandez_PRB18,Roszak_PRA19}. We had been motivated to look at correlations of the multiple measurements partially by a desire to address this question. However, as we show in \cite{Sakuldee_operations} the general structure of the relationship between correlators of multiple measurements on the qubit, and the coherence measured under dynamical decoupling, which underpins the results in this paper, is very general: it holds also for an environment described quantum-mechanically. In order to observe the back-action of the qubit on the environment or other quantum features of its dynamics one apparently has to involve both unitary operations and measurements applied to the qubit, in agreement with proposals from \cite{Fink_arXiv14,Bethke_arXiv19}.

\section*{Acknowledgements}
We thank Piotr Sza{\'n}kowski for a careful reading of the manuscript and discussions and Damian Kwiatkowski for discussions concerning the relation of this work to the results of \cite{Laraoui_NC13}. 
This work was supported by funds from Polish National Science Center (NCN), Grant No.~2015/19/B/ST3/03152.

\bibliographystyle{apsrev4-1}
\bibliography{correlations_of_measurements_classical_noise_PRA_revise_I.bbl}
%\bibliography{../../../../refs_quant,../../../../refs_entanglement,../../../../refs_ddns_magnetometry,../../../../refs_decoherence,../../../../refs_Si,multiplemeasurement,../../../../refs_measurement,multiplemeasurement}

\end{document}